\documentclass[11pt, prd, aps, showpacs, nofootinbib]{revtex4-1}

\usepackage{amsmath,amssymb}
\usepackage{graphicx}
\DeclareGraphicsExtensions{.pdf,.png,.jpg}

\newcommand{\be}{\begin{equation}}
\newcommand{\ee}{\end{equation}}
\newcommand{\bea}{\begin{eqnarray}}
\newcommand{\eea}{\end{eqnarray}}

\newcommand{\Regensburg}{Universit\"at Regensburg, Fakult\"at f\"ur Physik, D-93040, Regensburg, Germany\\
Istituto Nazionale di Fisica Nucleare, Sezione di Roma Tre, I-00146 Rome, Italy}
\newcommand{\RomatreINFN}{Istituto Nazionale di Fisica Nucleare, Sezione di Roma Tre, I-00146 Rome, Italy}

\begin{document}

\title{\Large Ratios of the hadronic contributions to the lepton $g - 2$\\[2mm] from Lattice QCD+QED simulations}

\author{\large D.~Giusti} \affiliation{\Regensburg}
\author{\large S.~Simula} \affiliation{\RomatreINFN}
\author{\large for the Extended Twisted Mass Collaboration} \noaffiliation

\begin{abstract}
\begin{center}
\vspace{0.25cm}
\includegraphics[draft=false,width=.25\linewidth]{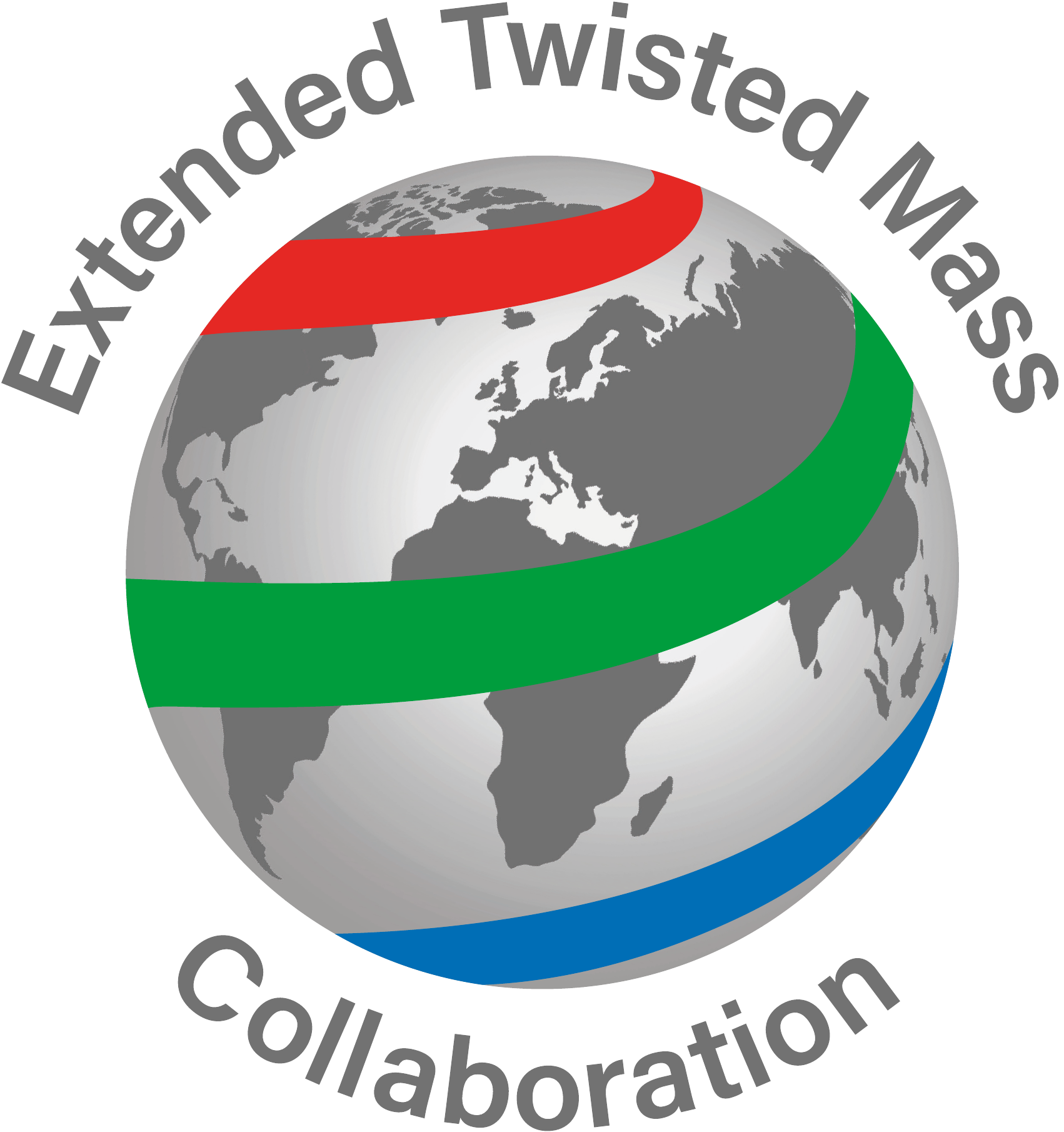}
\vspace{0.25cm}
\end{center}
The ratios among the leading-order (LO) hadronic vacuum polarization (HVP) contributions to the anomalous magnetic moments of electron, muon and $\tau$-lepton, $a_{\ell = e, \mu, \tau}^{\rm HVP, LO}$, are computed using lattice QCD+QED simulations. 
The results include the effects at order ${\cal{O}}(\alpha_{em}^2)$ as well as the electromagnetic and strong isospin-breaking corrections at orders ${\cal{O}}(\alpha_{em}^3)$ and ${\cal{O}}(\alpha_{em}^2 (m_u - m_d))$, respectively, where $(m_u - m_d)$ is the $u$- and $d$-quark mass difference.
We employ the gauge configurations generated by the Extended Twisted Mass Collaboration with $N_f = 2+1+1$ dynamical quarks at three values of the lattice spacing ($a \simeq 0.062, 0.082, 0.089$ fm) with pion masses in the range $\simeq 210 - 450$ MeV.
The calculations are based on the quark-connected contributions to the HVP in the quenched-QED approximation, which neglects the charges of the sea quarks.
The quark-disconnected terms are estimated from results available in the literature.
We show that in the case of the electron-muon ratio the hadronic uncertainties in the numerator and in the denominator largely cancel out, while in the cases of the electron-$\tau$ and muon-$\tau$ ratios such a cancellation does not occur. 
For the electron-muon ratio we get $R_{e / \mu } \equiv (m_\mu / m_e)^2 (a_e^{\rm HVP, LO} / a_\mu^{\rm HVP, LO}) = 1.1456~(83)$ with an uncertainty of $\simeq 0.7 \%$.
Our result, which represents an accurate Standard Model (SM) prediction, agrees very well with the estimate obtained using the results of dispersive analyses of the experimental $e^+ e^- \to$ hadrons data.
Instead, it differs by $\simeq 2.7$ standard deviations from the value expected from present electron and muon ($g - 2$) experiments after subtraction of the current estimates of the QED, electro-weak, hadronic light-by-light and higher-order HVP contributions, namely $R_{e / \mu} = 0.575~(213)$.
An improvement of the precision of both the experiment and the QED contribution to the electron ($g - 2$) by a factor of $\simeq 2$ could be sufficient to reach a tension with our SM value of the ratio $R_{e / \mu }$ at a significance level of $\simeq 5$ standard deviations.
\end{abstract}

\maketitle

\newpage

\section{Introduction}
\label{sec:intro}

Since many years a long standing deviation between experiment and theory persists for the anomalous magnetic moment of the muon, $a_\mu \equiv (g_\mu - 2) / 2$.
The E821 experiment~\cite{Bennett:2006fi,PDG} at Brookhaven National Lab  currently provides the most accurate measurement of $a_\mu$
\be
     a_\mu^{exp} = 11~659~209.1 ~ (5.4) ~ (3.3) ~ [6.3] \cdot 10^{-10} ~ ,
     \label{eq:amu_exp}
\ee
where the first error is statistical, the second one systematic and the third error in brackets is the sum in quadrature corresponding to a final accuracy of $0.54$ ppm.
An improvement of the uncertainty by a factor of four is in progress thanks to the experiment E989 at FermiLab~\cite{Logashenko:2015xab,Grange:2015fou} (and later to the experiment E34 at J-PARC~\cite{Abe:2019thb}). 
First results from E989 are expected in 2020.

On the theoretical side the present accuracy of the Standard Model (SM) prediction is at a similar level, $0.53$ ppm~\cite{PDG}.
According to the most recent determinations of the hadronic contributions to $a_\mu$, obtained using dispersive analyses of the experimentally measured $e^+ e^- \to$ hadrons data~\cite{Davier:2019can,Keshavarzi:2019abf}, the muon anomaly, i.e.~the difference between $a_\mu^{exp}$ and $a_\mu^{SM}$, is given by
\bea
    a_\mu^{exp}  - a_\mu^{SM} & = & 26.0 ~ (6.3)_{\rm exp} ~ (4.8)_{\rm th} ~ [7.9] ~ \cdot 10^{-10} ~ \qquad \mbox{\cite{Davier:2019can}} ~ , ~ \nonumber \\
                                                 & = & 28.0 ~ (6.3)_{\rm exp} ~ (3.8)_{\rm th} ~ [7.4] ~ \cdot 10^{-10} ~ \qquad \mbox{\cite{Keshavarzi:2019abf}} ~ ,
    \label{eq:anomaly_muon}
\eea
where the first error comes from experiment, the second one from theory and the third one is the sum in quadrature corresponding respectively to a final discrepancy of $\simeq 3.3$~\cite{Davier:2019can} and $\simeq 3.8$~\cite{Keshavarzi:2019abf} standard deviations.
Other estimates of the hadronic contributions to $a_\mu$, based always on the analysis of $e^+ e^- \to$ hadrons data, provide similar discrepancies (see, e.g., Ref.~\cite{Jegerlehner:2017lbd}).

A new interesting deviation occurs in the case of  the anomalous magnetic moment of the electron $a_e$, which has been measured at the very high level of accuracy of $0.24$ ppb~\cite{Hanneke:2008tm,Hanneke:2010au}
\be
     a_e^{exp} = 11~596~521~807~3 ~ [28] \cdot 10^{-14} ~ .
     \label{eq:ae_exp}
\ee
Thanks to a precise recent determination of the fine structure constant $\alpha_{em}^{-1} = 137.035~999~046 ~ (27)$ from Ref.~\cite{Parker:2018vye}, the SM prediction for $a_e$ corresponds to an electron anomaly equal to
\be
    a_e^{exp} - a_e^{SM} = -89 ~ (28)_{\rm exp} ~ (23)_{\rm th} ~ [36] ~ \cdot 10^{-14} \qquad \mbox{\cite{Keshavarzi:2019abf,Aoyama:2019ryr}} ~ ,
    \label{eq:anomaly_electron}
\ee
where the theory error is dominated by the uncertainty on $\alpha_{em}$ and the final error corresponds to a discrepancy of $\simeq 2.5$ standard deviations.
Note that the electron anomaly (\ref{eq:anomaly_electron}) is opposite in sign with respect to the muon anomaly (\ref{eq:anomaly_muon}).

On the contrary no direct measurement of the anomalous magnetic moment of the third charged lepton of the SM, the $\tau$ lepton, is available due to its short lifetime.
Only limits have been set in an indirect way by the DELPHI Collaboration~\cite{Abdallah:2003xd} to be $-0.052 < a_\tau^{exp} < 0.013$ at the $95 \%$ confidence level.
The precision is quite poor even with respect to the one-loop QED contribution $\alpha_{em} / 2 \pi \sim {\cal{O}}(10^{-3})$~\cite{Schwinger:1948zz}.
Nevertheless, the quantity $a_\tau$ is considered to be the best candidate for finding physics beyond the SM, since for a large class of theories the contribution of new physics to the lepton anomalous magnetic moments is proportional to the squared lepton mass\footnote{In this respect note that the absolute value of the electron anomaly (\ref{eq:anomaly_electron}) is larger by an order of magnitude than the value $\approx 6.5 \cdot 10^{-14}$ expected naively from the muon anomaly (\ref{eq:anomaly_muon}) and the lepton-mass scaling $m_e^2 / m_\mu^2$ (see, e.g., Refs.~\cite{Giudice:2012ms,Davoudiasl:2018fbb,Crivellin:2018qmi}).}.

For the three leptons the SM prediction of their anomalous magnetic moments is given by the sum of three contributions
\be
    a_\ell^{SM} = a_\ell^{QED} + a_\ell^{EW} + a_\ell^{had} ~ \qquad \qquad (\ell = e, \mu, \tau) ~ ,
    \label{eq:aell}
\ee
where $a_\ell^{QED}$ is the QED term known up to five loops~\cite{Aoyama:2019ryr}, $a_\ell^{EW}$ represents the electroweak (EW) corrections known up to two loops~\cite{Jegerlehner:2017zsb,Gnendiger:2013pva,Eidelman:2007sb} and $a_\ell^{had}$ is the hadronic term, which includes the hadronic vacuum polarization (HVP) and the light-by-light (LBL) contributions
\be
    a_\ell^{had} = a_\ell^{HVP} + a_\ell^{LBL} ~ .
    \label{eq:aell_hadron}
\ee

Precise determinations of $a_\ell^{HVP}$ come from dispersion relations and the experimentally measured $e^+ e^- \to$ hadrons data, while $a_\ell^{LBL}$ have been estimated through phenomenological hadronic models and by dispersive approaches (see Ref.~\cite{Colangelo:2017fiz} and therein).
Both quantities are non-perturbative and, therefore, they should be calculated from first principles, i.e.~by means of lattice QCD+QED simulations. 

During the last years a tremendous effort has been put to obtain accurate determinations of both $a_\mu^{HVP}$ and $a_\mu^{LBL}$ by various lattice collaborations.
The present status and the perspectives of the lattice calculations of both $a_\mu^{HVP}$ and $a_\mu^{LBL}$ have been discussed in a series of workshops of the {\it Muon (g-2) Theory Initiative}~\cite{TIgm2}, which has produced the recent White Paper of Ref.~\cite{Aoyama:2020ynm}.
The main outcome is that for $a_\mu^{HVP}$ the overall lattice precision is not yet competitive with respect to the one of the dispersive results, while recent lattice estimates of the LBL term are consistent with the phenomenological and dispersive findings within the current level of precision (see for details Ref.~\cite{Aoyama:2020ynm}).
Recently the BMW Collaboration~\cite{Borsanyi:2020mff} claims to have reached a precision for $a_\mu^{HVP}$ similar to the one of the dispersive approaches, although getting a significant discrepancy for the central values (see also Refs.~\cite{Crivellin:2020zul,Keshavarzi:2020bfy} for implications on global fits to EW precision observables).  

As far as the electron and the $\tau$-lepton are concerned, only two lattice estimates of the HVP contribution from Refs.~\cite{Burger:2015oya,Borsanyi:2017zdw} exist to date.

The aim of this work is to present a lattice determination of the ratios of the leading-order (LO) HVP contributions to the lepton anomalous magnetic moments $a_e$, $a_\mu$ and $a_\tau$, obtained using the same hadronic input determined by the lattice QCD+QED simulations of Refs.~\cite{Giusti:2017jof,Giusti:2018mdh,Giusti:2019xct}, where the gauge configurations generated by the Extended Twisted Mass Collaboration (ETMC) with $N_f = 2+1+1$ dynamical quarks at three values of the lattice spacing ($a \simeq 0.062, 0.082, 0.089$ fm) with pion masses in the range $\simeq 210 - 450$ MeV~\cite{Baron:2010bv,Baron:2011sf} were adopted.
The lattice framework and details of the simulations are summarized in Appendix~\ref{sec:appA}.

Our simulations include the effects at order ${\cal{O}}(\alpha_{em}^2)$ as well as the electromagnetic (em) and strong isospin-breaking (IB) corrections at orders ${\cal{O}}(\alpha_{em}^3)$ and ${\cal{O}}(\alpha_{em}^2 (m_u - m_d))$, respectively, where $(m_u - m_d)$ is the $u$- and $d$-quark mass difference. 
The calculations are based on quark-connected contributions to the HVP in the quenched QED (qQED) approximation, which neglects the charges of the sea quarks.
The quark-disconnected terms can be estimated from results available in the literature (see Refs.~\cite{Blum:2015you,Borsanyi:2017zdw,Blum:2018mom,Borsanyi:2020mff}).
The ETMC results for $a_e^{\rm HVP, LO}$, $a_\mu^{\rm HVP, LO}$ and $a_\tau^{\rm HVP, LO}$ at the physical point have been presented already in Refs.~\cite{Giusti:2019hoy,Giusti:2019hkz} and exhibit uncertainties at the level of $\simeq 2 \div 2.4 \%$.

We stress that the hadronic quantities $a_\ell^{\rm HVP, LO}$ for $\ell = e, \mu, \tau$ share the same hadronic input and differ only in the leptonic kinematical kernel.
We show that among the various ratios of $a_\ell^{\rm HVP, LO}$ for different leptons the electron-muon ratio play a special role, since in this case the hadronic uncertainties in the numerator and in the denominator are strongly correlated and largely cancel out.
The same does not occur in the case of the electron-$\tau$ and muon-$\tau$ ratios, where the numerator and the denominator turn out to be almost uncorrelated.

For the electron-muon ratio we get\footnote{In Eq.~(\ref{eq:ratio_mue}) we have introduced the factor $(m_\mu / m_e)^2$ so that the ratio $R_{e / \mu}$ differs from unity only due to the curvature and higher-order Mellin-Barnes moments (and their derivatives) of the HVP function at vanishing photon virtuality~\cite{deRafael:2014gxa}. For the mass ratio $m_\mu / m_e$ we adopt the CODATA value $m_\mu / m_e = 206.7682831 ~ (47)$ from Ref.~\cite{Mohr:2015ccw}.}
\be
    R_{e / \mu} \equiv \left( \frac{m_\mu}{m_e} \right)^2 ~ \frac{a_e^{\rm HVP, LO}}{a_\mu^{\rm HVP, LO}} = 1.1456 ~ (83) ~  , 
    \label{eq:ratio_mue}
\ee
where the error includes both statistical and systematic uncertainties and corresponds to a hadronic uncertainty of $\simeq 0.7 \%$, i.e.~a factor $\approx 4$ better than the individual precisions of the numerator and the denominator.

Our result (\ref{eq:ratio_mue}), which represents an accurate SM prediction, agrees very well with the one corresponding to the results of the dispersive analyses of $e^+ e^- \to$ hadrons data carried out recently in Ref.~\cite{Keshavarzi:2019abf}, namely $a_e^{\rm HVP, LO}(e^+ e^-) = 186.08~(0.66) \cdot 10^{-14}$ and $a_\mu^{\rm HVP, LO}(e^+ e^-) = 692.78~(2.42) \cdot 10^{-10}$ leading to $R_{e / \mu}^{e^+ e^-} = 1.1483 ~ (41)_e ~ (40)_\mu ~ [57]$, where the first and second errors are related to the electron and muon contributions separately, while the third error is their sum in quadrature, i.e.~without taking into account correlations between the numerator and the denominator.

Let us now introduce the following HVP quantities $\overline{a}_\ell^{\rm HVP, LO}$ defined as
\be
     \overline{a}_\ell^{\rm HVP, LO} \equiv a_\ell^{exp} - a_\ell^{QED} - a_\ell^{EW} - a_\ell^{LBL} - a_\ell^{HVP, HO} ~ ,
     \label{eq:aell_exp-QED}
\ee
where $a_\ell^{HVP, HO}$ denotes the higher-order HVP corrections due to multiple insertions of leptonic and hadronic loops.
In the case of the electron and the muon, adopting for the quantities in the r.h.s.~of Eq.~(\ref{eq:aell_exp-QED}) the same inputs from Ref.~\cite{Keshavarzi:2019abf} leading to the anomalies (\ref{eq:anomaly_electron}) and (\ref{eq:anomaly_muon}), one gets
\bea
      \label{eq:ae_exp-QED}
      \overline{a}_e^{\rm HVP, LO} & = & 97 ~ (28)_{\rm exp} ~ (23)_{\rm th} ~ [36] \cdot 10^{-14} ~ , \\[2mm]
      \label{eq:amu_exp-QED}
      \overline{a}_\mu^{\rm HVP, LO} & = & 720.8 ~ (6.3)_{\rm exp} ~ (2.9)_{\rm th} ~ [6.9] \cdot 10^{-10} ~ ,
\eea
where the theoretical uncertainties come mainly from the QED contribution for the electron and from the hadronic LBL term for the muon. 

The results (\ref{eq:ae_exp-QED}-\ref{eq:amu_exp-QED}) imply a value for the electron-muon ratio $R_{e / \mu}$ (which for sake of simplicity will be referred to as the ``exp - QED" value) equal to
\be
     R_{e / \mu}^{\rm exp - QED} \equiv \left( \frac{m_\mu}{m_e} \right)^2 ~ \frac{\overline{a}_e^{\rm HVP, LO}}{\overline{a}_\mu^{\rm HVP, LO}} = 0.575 ~ (213)_e ~ (6)_\mu ~ [213] ~ ,
     \label{eq:ratio_mue_exp-QED}
\ee
which differs from our lattice result (\ref{eq:ratio_mue}) by $\simeq 2.7$ standard deviations corresponding to a tension governed mainly by the one of the electron anomaly (\ref{eq:ae_exp-QED}).
An improvement by a factor of $\simeq 2$ in the precision of both the experiment and the QED contribution for the electron might be enough to reach a significance level of $\simeq 5$ standard deviations from our SM value (\ref{eq:ratio_mue}) as well as for the electron anomaly itself.

The plan of the paper is as follows.

In Section~\ref{sec:kernels} we briefly summarize the way we calculate the LO HVP terms $a_\ell^{\rm HVP, LO}$ and present also an explicit comparison among the kinematical kernels for the three leptons $\ell = e, \mu, \tau$.

In Section~\ref{sec:Vt} we describe our results obtained using the same hadronic input shared by all the three leptons, i.e.~the vector correlator $V(t)$, adopting the ETMC gauge ensembles described in Appendix~\ref{sec:appA}. 
We define also the electron-muon ratio $R_{e / \mu}$ and present our calculations of the light-quark contribution in Section~\ref{sec:light}. 
By using the ``dual + $\pi \pi$'' representation of the vector correlator $V^{ud}(t)$, developed in Ref.~\cite{Giusti:2018mdh} and described in Appendix~\ref{sec:appB}, we correct our data for finite-volume effects.
Then, by adopting three different strategies we perform the chiral extrapolation to the physical pion mass and also to the continuum limit.
The remaining contributions to $R_{e / \mu}$ are evaluated in Section~\ref{sec:Remu_other}. 

In Section~\ref{sec:results} we present our determinations of the three ratios $R_{e / \mu}$, $R_{e / \tau}$ and $R_{\mu / \tau}$, extrapolated to the physical pion mass and to the continuum and infinite volume limits. 
We show that our results for the three ratios agree well with those corresponding to the recent analyses of $e^+ e^- \to$ hadrons data from Ref.~\cite{Keshavarzi:2019abf} as well as with an estimate of $R_{e / \mu}$, which we derive from the BMW results of Ref.~\cite{Borsanyi:2017zdw}.

Section~\ref{sec:conclusions} collects our conclusions and perspectives.

\section{The LO HVP contribution to the lepton $a_\ell$}
\label{sec:kernels}

The LO HVP contribution $a_\ell^{\rm HVP, LO}$ to the lepton anomalous magnetic moment ($\ell = e, \mu, \tau$) is related to the Euclidean HVP function $\Pi(Q^2)$ by~\cite{Lautrup:1971jf,deRafael:1993za,Blum:2002ii}
 \be
      a_\ell^{\rm HVP, LO} = 4 \alpha_{em}^2 \int_0^\infty dQ^2 f_\ell(Q^2) \left[ \Pi(Q^2) -  \Pi(0) \right] ~ ,
      \label{eq:aell_HVP}
 \ee
where $Q$ is the Euclidean four-momentum and the leptonic kernel $f_\ell(Q^2)$ is given by
\be
     f_\ell(Q^2) = \frac{1}{m_\ell^2} ~ \frac{1}{\omega} ~ \frac{1}{\sqrt{4 + \omega^2}} ~ \left( \frac{\sqrt{4 + \omega^2} - \omega}{\sqrt{4 + \omega^2} + \omega} \right)^2
     \label{eq:fell_Q2}
 \ee
with $m_\ell$ being the lepton mass and $\omega \equiv Q / m_\ell$.

The HVP form factor $\Pi(Q^2)$ contains the non-perturbative hadronic effects and it is defined through the HVP tensor as
 \be
      \Pi_{\mu \nu}(Q) \equiv \int d^4x ~ e^{iQ \cdot x} \langle J_\mu(x) J_\nu(0) \rangle = ( \delta_{\mu \nu} Q^2 - Q_\mu Q_\nu ) \Pi(Q^2)
      \label{eq:HVPtensor}
 \ee
where
 \be
     J_\mu(x) \equiv \sum_{f = u, d, s, c, ...} q_f ~ \overline{\psi}_f(x) \gamma_\mu \psi_f(x) 
     \label{eq:Jmu}
 \ee 
is the em current operator with $q_f$ being the electric charge of the quark with flavor $f$ in units of the electron charge $e$, while $\langle ... \rangle$ means the average of the $T$-product of the two em currents over gluon and quark fields.
In Eq.~(\ref{eq:aell_HVP}) the {\it subtracted} HVP function $\Pi_R(Q^2) \equiv \Pi(Q^2) - \Pi(0)$ appears in order to guarantee that the em coupling $\alpha_{em}$ is the experimental one in the Thomson limit (i.e.~$Q^2 << m_e^2$).

In this work we adopt the time-momentum representation of Ref.~\cite{Bernecker:2011gh}, in which the HVP function $\Pi_R(Q^2)$ is expressed as
 \be
      \Pi_R(Q^2) = \Pi(Q^2) -  \Pi(0) = 2 \int_0^\infty dt ~ V(t) \left[ \frac{\mbox{cos}(Qt) -1}{Q^2} + \frac{1}{2} t^2 \right] ~ ,
      \label{eq:cos}
 \ee
where $V(t)$ is the vector current-current Euclidean correlator defined as
 \be
     V(t) \equiv - \frac{1}{3} \sum_{i=1,2,3} \int d\vec{x} ~ \langle J_i(\vec{x}, t) J_i(0) \rangle 
     \label{eq:VVt}
 \ee
 and $t$ is the Euclidean time distance.
Thus, the LO HVP contribution $a_\ell^{HVP,\,LO}$ reads as
 \be
      a_\ell^{\rm HVP, LO} = 4 \alpha_{em}^2 \int_0^\infty dt ~ K_\ell(t) V(t) ~ ,
      \label{eq:aell_t}
 \ee
 where
 \bea
      K_\ell(t) & \equiv & 2 \int_0^\infty dQ^2 ~ f_\ell(Q^2) \left[ \frac{\mbox{cos}(Qt) -1}{Q^2} + \frac{1}{2} t^2 \right] \nonumber \\[2mm] 
                    & = & t^2 \int_0^1 dx ~ (1 - x) \left[ 1 - j_0^2\left( \frac{m_\ell t}{2} \frac{x}{\sqrt{1-x}} \right) \right]
      \label{eq:kernel_ell}
 \eea
with $j_0(y)$ being the spherical Bessel function $j_0(y) = \mbox{sin}(y) / y$ and $Q^2 \equiv m_\ell^2 x^2 / (1-x)$.
 
\begin{figure}[htb!]
\begin{center}
\includegraphics[scale=0.75]{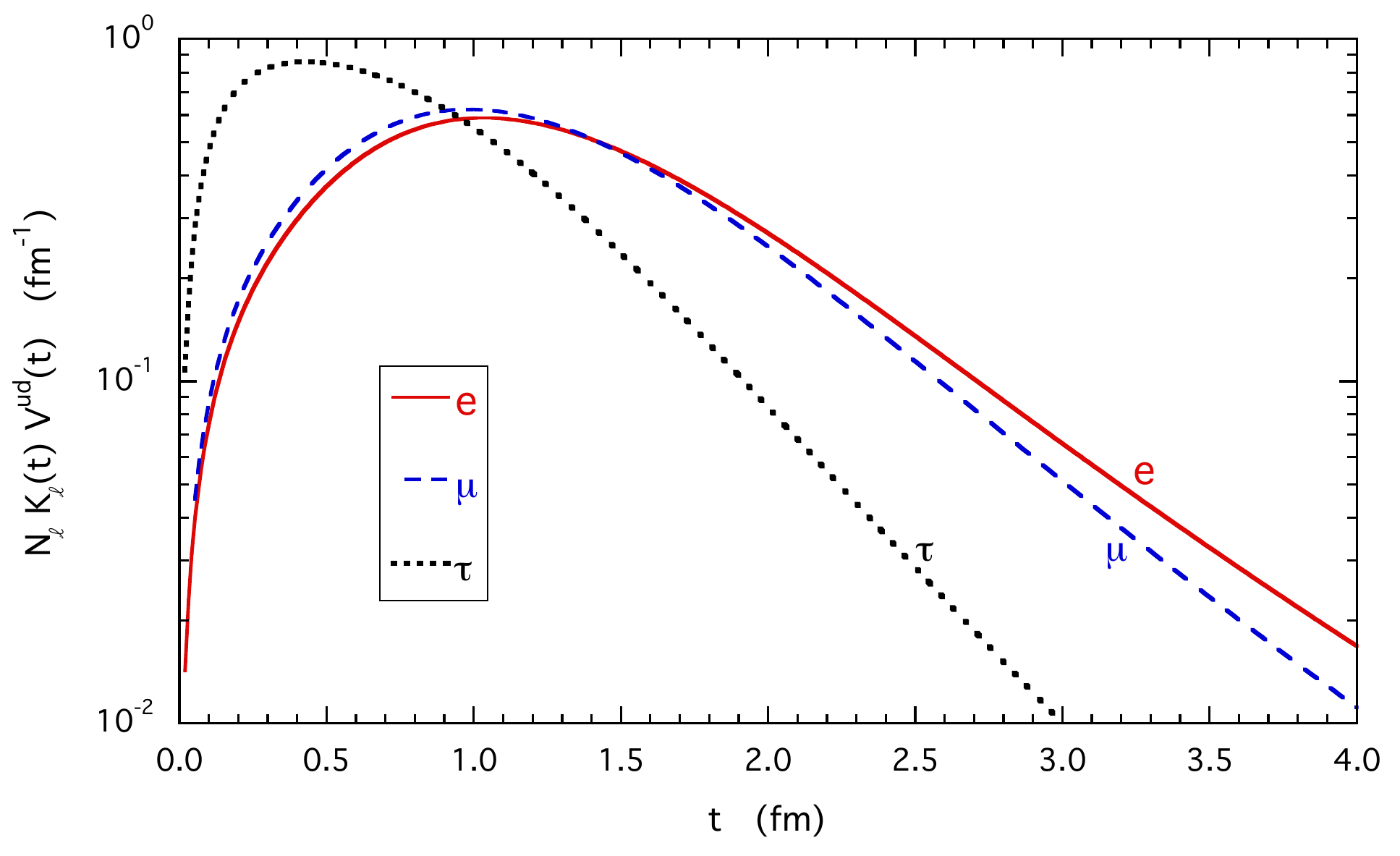}
\end{center}
\vspace{-1.0cm}
\caption{\it \small Comparison of the normalized quantities $N_\ell K_\ell(t) V^{ud}(t)$ for $\ell = \{ e, \mu, \tau \}$ versus the time distance $t$. The three kernels $K_\ell(t)$ are given by Eq.~(\ref{eq:kernel_ell}). The hadronic quantity $V^{ud}(t)$ is the light-quark (connected) contribution to the vector current-current correlator (\ref{eq:VVt}), as determined at the physical point in Ref.~\cite{Giusti:2018mdh} (see later Section~\ref{sec:light} and Appendices~\ref{sec:appA} and~\ref{sec:appB}). The constants $N_\ell$ are introduced in order to guarantee the common normalization condition $N_\ell \int_0^\infty dt K_\ell(t) V^{ud}(t)= 1$ for all leptons, while the uncertainties of $V^{ud}(t)$ are not shown.}
\label{fig:kernels}
\end{figure}

The LO HVP contributions $a_\ell^{\rm HVP, LO}$, given by Eq.~(\ref{eq:aell_t}), have in common the hadronic input $V(t)$ and differs only in the kernels $K_\ell(t)$, which weigh different temporal regions differently according to the lepton masses involved.
For purposes of illustration let us use for the hadronic input $V(t)$ its light-quark (connected) component $V^{ud}(t)$ determined at the physical point in Ref.~\cite{Giusti:2018mdh} (see later Section~\ref{sec:light} and Appendices~\ref{sec:appA} and~\ref{sec:appB}).
In Fig.~\ref{fig:kernels} the $t$-dependencies of the quantities $N_\ell K_\ell(t) V^{ud}(t)$ are compared for the three cases $\ell = \{ e, \mu, \tau \}$.
The constants $N_\ell$ are introduced in order to guarantee the common normalization condition $N_\ell \int_0^\infty dt K_\ell(t) V^{ud}(t) = 1$ for all leptons, while the uncertainties on $V^{ud}(t)$ are not shown.

It can be seen that the time distances relevant for the integration in the r.h.s.~of Eq.~(\ref{eq:aell_t}) are quite similar in the case of the electron and the muon. 
Instead, in the case of the $\tau$-lepton the impact of the short and intermediate time distances up to $1 \div 1.5$ fm is enhanced, while the role of the large time distances is reduced.
We expect therefore that by considering ratios $a_\ell^{\rm HVP, LO} / a_{\ell^\prime}^{\rm HVP, LO}$ for different leptons the correlation between the numerator and the denominator should be significant mainly in the case of the electron-muon ratio.

\section{The hadronic input $V(t)$}
\label{sec:Vt}

Thanks to recent progress in lattice QCD+QED simulations the vector current-current correlator (\ref{eq:VVt}) is nowadays calculated including both strong and em IB corrections, related to the mass difference $(m_d - m_u)$ between $u$- and $d$-quarks and to the em interactions among quarks.
Since both $(m_d - m_u) / \Lambda_{QCD}$ and $\alpha_{em}$ are {\it small} parameters of order ${\cal{O}}(1 \%)$, an expansion of the path integral in powers of the two parameters has been developed in Refs.~\cite{deDivitiis:2011eh,deDivitiis:2013xla}.
Thus, the vector correlator $V(t)$ can be split into two contributions
\be
    V(t) = V^{\rm isoQCD}(t) + \delta V^{\rm IB}(t) ~ ,
    \label{eq:VVt_RM123}
\ee
where $V^{\rm isoQCD}(t)$ corresponds to the contribution of isosymmetric QCD only (i.e., $m_u = m_d$ and $\alpha_{em} = 0$), while $\delta V^{\rm IB}(t)$ includes the contributions at first order ${\cal{O}}((m_d - m_u) / \Lambda_{QCD})$ and ${\cal{O}}(\alpha_{em})$.
Terms at higher orders are sub-leading and they can be safely neglected even for a permil-precision calculation of the HVP term $a_\ell^{\rm HVP, LO}$.

It should be stressed that the separation given in Eq.~(\ref{eq:VVt_RM123}) requires a prescription (see Section II of Ref.~\cite{DiCarlo:2019thl} for an exhaustive discussion), which means that both $V^{\rm isoQCD}(t)$ and $\delta V^{\rm IB}(t)$ are prescription dependent. 
Only the complete correlator $V(t)$ (and correspondingly the HVP term $a_\ell^{\rm HVP, LO}$) is prescription free. 
In this work we follow Refs.~\cite{Giusti:2018mdh,Giusti:2017jof,Giusti:2019xct} and adopt the Gasser-Rusetsky-Scimemi  prescription~\cite{Gasser:2003hk}, in which the renormalized quark masses and strong coupling, evaluated in the $\rm \overline{MS}$ scheme at a renormalization scale of 2 GeV, are equal in the full QCD+QED and isosymmetric QCD theories.

Since all quark flavors contribute to the em current (\ref{eq:Jmu}), both $V^{\rm isoQCD}(t)$ and $\delta V^{\rm IB}(t)$ can be written as
\bea
      \label{eq:V0_t}
      V^{\rm isoQCD}(t) & = & V^{ud}(t) + V^s(t) + V^c(t) + V^{disc}(t) ~ , ~ \\[2mm]
      \label{eq:deltaV_t}
      \delta V^{\rm IB}(t) & = & \delta V^{ud}(t) + \delta V^s(t) + \delta V^c(t) + \delta V^{disc}(t) ~ ,
\eea
where the first three terms in the r.h.s.~correspond to the contribution of light, strange and charm quark flavor separately (quark-connected contractions), while the fourth term represents the contribution of quark-disconnected diagrams.
We have not included any contribution from the bottom quark, since it is sub-leading with respect even to a permil-precision level\footnote{In the case of the muon the bottom-quark LO HVP contribution $a_\mu^{\rm HVP, LO}(b)$ has been found to be equal to $0.271\,(37) \cdot 10^{-10}$ in Ref.~\cite{Colquhoun:2014ica} in agreement with the perturbative QCD estimate $0.29\,(1) \cdot 10^{-10}$~\cite{Bodenstein:2011qy}.}.

Correspondingly, from Eqs.~(\ref{eq:VVt_RM123}-\ref{eq:deltaV_t}) one has
\be
    a_\ell^{\rm HVP, LO} = a_\ell^{\rm HVP, LO}(\rm isoQCD) + a_\ell^{\rm HVP, LO}(\rm IB)
    \label{eq:ajHVP_RM123}
\ee
with
\bea
      \label{eq:ajHVP_LO}
      a_\ell^{\rm HVP, LO}(\rm isoQCD) & = & a_\ell^{\rm HVP, LO}(ud) + a_\ell^{\rm HVP, LO}(s) + a_\ell^{\rm HVP, LO}(c) + a_\ell^{\rm HVP, LO}(disc) ~ , ~  \\[4mm]
      \label{eq:ajHVP_delta}
      a_\ell^{\rm HVP, LO}(\rm IB) & = & \delta a_\ell^{\rm HVP, LO}(ud) + \delta a_\ell^{\rm HVP, LO}(s) + \delta a_\ell^{\rm HVP, LO}(c) + \delta a_\ell^{\rm HVP, LO}(disc) ~ , \qquad
\eea
where all the terms in $a_\ell^{\rm HVP, LO}(\rm isoQCD)$ are of order ${\cal{O}}(\alpha_{em}^2)$, while those in $a_\ell^{\rm HVP, LO}(\rm IB)$ contain IB contributions at orders ${\cal{O}}(\alpha_{em}^2 (m_d - m_u)/ \Lambda_{QCD})$ and ${\cal{O}}(\alpha_{em}^3)$.

We start by considering the electron-muon ratio $R_{e / \mu}$ given by Eq.~(\ref{eq:ratio_mue}).
Since the (connected) light-quark contribution $a_\ell^{\rm HVP, LO}(ud)$ represents almost $90 \%$ of the total LO HVP term $a_\ell^{HVP,\,LO}$, we rewrite the ratio $R_{e / \mu}$ in the following form
 \be
    R_{e / \mu} \equiv R_{e / \mu}^{ud} \cdot \widetilde{R}_{e / \mu} ~ ,
    \label{eq:Remu}
 \ee
where
 \be
    R_{e / \mu}^{ud} \equiv \left( \frac{m_\mu}{m_e} \right)^2 ~ \frac{a_e^{\rm HVP, LO}(ud)}{a_\mu^{\rm HVP, LO}(ud)}
    \label{eq:Remu_ud}
 \ee
and
 \be
    \widetilde{R}_{e / \mu} \equiv \frac{1 + \sum_{j = s, c, IB, disc} ~ \frac{a_e^{\rm HVP, LO}(j)}{a_e^{\rm HVP, LO}(ud)}}{1 + \sum_{j = s, c, IB, disc} ~  \frac{a_\mu^{\rm HVP, LO}(j)}{a_\mu^{\rm HVP, LO}(ud)}}
    \label{eq:Remu_tilde}
 \ee

In the next two subsections we address separately the determination of $R_{e / \mu}^{ud}$ and $\widetilde{R}_{e / \mu}$.

\subsection{Light-quark contribution $R_{e / \mu}^{ud}$}
\label{sec:light}

The results obtained for the ratio $R_{e / \mu}^{ud}$ adopting the $N_f = 2 + 1 + 1$ ETMC gauge ensembles of Appendix~\ref{sec:appA} are shown in Fig.~\ref{fig:ratioud_ETMC} as empty markers versus the simulated pion mass $M_\pi$.
The errors include (in quadrature) both statistical and systematic uncertainties according to the bootstrap samples generated for the input parameters of the quark mass analysis of Ref.~\cite{Carrasco:2014cwa}.
They are described in Appendix~\ref{sec:appA} and have been used in all our works on the muon HVP terms~\cite{Giusti:2017jof,Giusti:2018mdh,Giusti:2019xct}.
In the numerical simulations we have adopted a local version of the em current (\ref{eq:Jmu}), which requires in our lattice setup a multiplicative renormalization.
The latter one however cancels out exactly in the ratio $R_{e / \mu}^{ud}$ (as well as also in $\widetilde{R}_{e / \mu}$).
\begin{figure}[htb!]
\begin{center}
\includegraphics[scale=0.80]{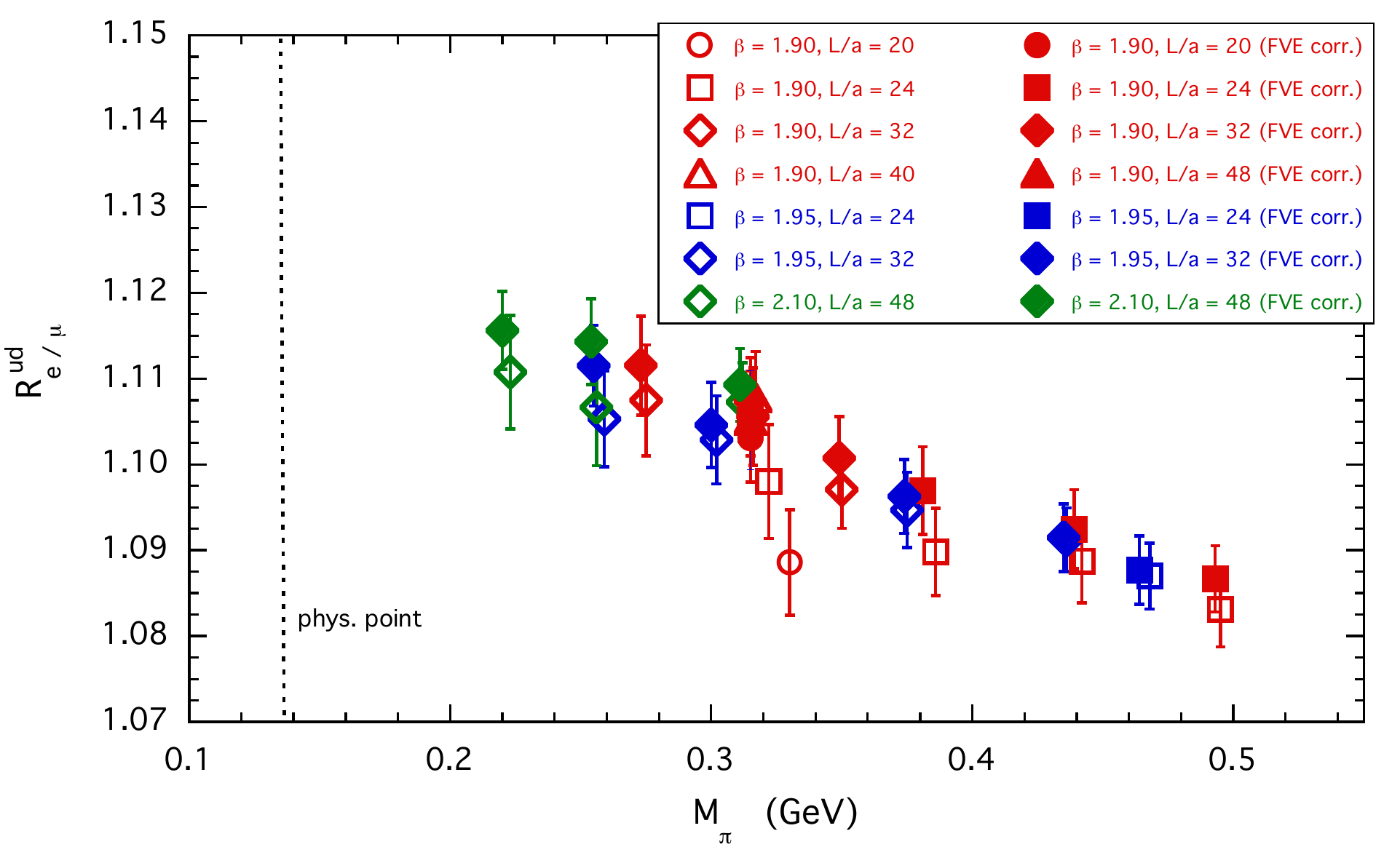}
\end{center}
\vspace{-1.0cm}
\caption{\it \small Results for the (connected) light-quark contribution to the electron-muon ratio, $R_{e / \mu}^{ud}$, versus the simulated pion mass $M_\pi$ for the $N_f = 2 + 1 + 1$ ETMC gauge ensembles of Appendix~\ref{sec:appA}. Empty markers correspond to the data computed at finite lattice size $L$, while full markers represent the ratio $R_{e / \mu}^{ud}(L \to \infty)$ corrected for FVEs according to Eq.~(\ref{eq:Rud_FVE}) evaluated using the results of Ref.~\cite{Giusti:2018mdh}. For each gauge ensemble the pion mass in the infinite volume limit is evaluated according to Ref.~\cite{Giusti:2018mdh}. Errors include (in quadrature) both statistical and systematic uncertainties according to the eight branches of the analyses described in Appendix~\ref{sec:appA}.}
\label{fig:ratioud_ETMC}
\end{figure}

Few comments are in order.
\begin{itemize} 
\item the precision of the data ranges from $\simeq 0.35 \%$ to $\simeq 0.6 \%$, i.e.~a reduction by a factor of at least $\simeq 4$ with respect to the precision of the individual HVP terms $a_\mu^{\rm HVP, LO}(ud)$ and $a_e^{\rm HVP, LO}(ud)$ achieved in Refs.~\cite{Giusti:2018mdh,Giusti:2019hkz}. This is clearly due to a significative correlation expected between the numerator and the denominator. Using the individual uncertainties we estimate the above correlation to be $\simeq 0.98$, i.e.~very close to $100 \%$;
\item the uncertainties of the data are mainly related to the statistical errors and to a lesser extent to the scale setting\footnote{By switching off the uncertainties of the scale setting in the bootstrap samples of Appendix~\ref{sec:appA} we get that the impact of the uncertainty on the scale setting does not exceed $\sim  15 \%$ of the errors of the calculated ratio $R_{e / \mu}^{ud}$ at the lightest simulated pion masses and reaches $\sim  40 \%$ only at the heaviest ones.};
\item finite volume effects (FVEs) are clearly visible in the case of the four gauge ensembles A40.XX (see Appendix~\ref{sec:appA}), which share the same pion mass and lattice spacing and differ only in the lattice size $L$; \item
the pion mass dependence is significative and the extrapolation to the physical pion mass requires a careful treatment, while discretization effects appear to be subleading.
\end{itemize}

In order to remove FVEs from the data we follow the approach of Ref.~\cite{Giusti:2018mdh}, where an analytic representation of the temporal dependence of $V^{ud}(t)$ was developed adopting the quark-hadron duality~\cite{SVZ} at short and intermediate time distances and the two-pion contribution in a finite box at large time distances~\cite{Luscher:1985dn,Luscher:1986pf,Luscher:1990ux,Luscher:1991cf,Lellouch:2000pv,Meyer:2011um,Francis:2013qna}.
A brief description of the analytic representation is illustrated in Appendix~\ref{sec:appB}. 
An accurate reproduction of the lattice data for $V^{ud}(t)$ was obtained for all the ETMC ensembles of Appendix~\ref{sec:appA} and the extrapolation to the infinite volume limit for the analytic representation of $V^{ud}(t)$ was achieved at each simulated pion mass and lattice spacing.

Using the analytic representation, the FVEs on $a_\mu^{\rm HVP, LO}(ud)$ were estimated in a non-perturbative way directly on the lattice~\cite{Giusti:2018mdh} and shown to differ significantly from the prediction of Chiral Perturbation Theory (ChPT) at next-to-leading order (NLO) up to values of $M_\pi L \approx 6$ (see also Refs.~\cite{DellaMorte:2017dyu,Giusti:2019dmu}).
Later FVEs on $a_\mu^{\rm HVP, LO}(ud)$ have been calculated within ChPT at NNLO~\cite{Aubin:2015rzx}. 
Still our findings differ from the NNLO predictions for values of $M_\pi L$ up to $\approx 5$.

The FVEs are subtracted from the data at finite volume using the following formula
 \be
    R_{e / \mu}^{ud}(L \to \infty) = R_{e / \mu}^{ud}(L) ~ \frac{a_e^{\rm HVP, LO}(ud; L \to \infty) / a_e^{\rm HVP, LO}(ud; L)}{a_\mu^{\rm HVP, LO}(ud; L \to \infty) / a_\mu^{\rm HVP, LO}(ud; L)} ~ ,
    \label{eq:Rud_FVE}
 \ee
where the two separate ratios $a_e^{\rm HVP, LO}(ud; L \to \infty) / a_e^{\rm HVP, LO}(ud; L)$ and $a_\mu^{\rm HVP, LO}(ud; L \to \infty) /$ $a_\mu^{\rm HVP, LO}(ud; L)$ are evaluated using the analytic representation of $V^{ud}(t)$.
The latter ones are strongly correlated so that the calculated correction due to FVEs on the ratio $R_{e / \mu}^{ud}(L)$ does not exceed $\simeq 1.3 \%$ with an uncertainty not larger than $\simeq 0.3 \%$.
The correlations between $R_{e / \mu}^{ud}(L)$ and the FVE correction, appearing in the r.h.s.~of Eq.~(\ref{eq:Rud_FVE}), are properly taken into account by means of our bootstrap procedure (see Appendix~\ref{sec:appA}). 
The data for $R_{e / \mu}^{ud}(L \to \infty)$ are shown in Fig.~\ref{fig:ratioud_ETMC} as full markers, while the values of both $R_{e / \mu}^{ud}(L)$ and $R_{e / \mu}^{ud}(L \to \infty)$ are given explicitly in Table~\ref{tab:data_ud} at the end of Appendix~\ref{sec:appA}.

The final steps are the extrapolations to the physical pion mass and to the continuum limit.
For evaluating the former one, which represents the dominant source of the systematic uncertainty, we adopt three strategies, which will be described in what follows.

In Ref.~\cite{Giusti:2018mdh} it was shown that for a proper chiral extrapolation of the ETMC data on $a_\mu^{\rm HVP, LO}(ud; L \to \infty)$ the effects of the chiral logs predicted by SU(3) ChPT at NLO and NNLO for the HVP form factor $\Pi_R^{ud}(Q^2)$ should be taken into account.
We point out that the chiral extrapolation of Ref.~\cite{Giusti:2018mdh} is only inspired by ChPT.
What we borrow from ChPT is the presence of chiral logs, i.e.~of non-analytic terms in the light-quark mass. 
These terms contribute to the pion mass dependence of $a_\ell^{\rm HVP, LO}(ud)$ regardless of the convergence properties of ChPT.
They might be expanded locally as powers of $M_\pi^2$, but in the case of our ETMC simulations, carried out for pion masses larger than $\sim 210$ MeV, it is unavoidable to use explicitly the chiral logs, otherwise any polynomial expansion would require too many terms to take into account the effects of the logs from the simulated pion masses down to the physical pion point.
The apparent linear behavior of the ETMC data shown in Fig.~\ref{fig:ratioud_ETMC} may be a consequence of the resummed higher-orders, which are not calculable using ChPT.

Therefore, we adopt the following Ansatz
 \be
    R_{e / \mu}^{ud}(L \to \infty) = \left[ \frac{m_\mu^2}{m_e^2} \left( \frac{a_e^{\rm HVP, LO}(ud)}{a_\mu^{\rm HVP, LO}(ud)} \right)^{ChPT} + A_0 + A_1 M_\pi^2 \right] 
                                                    \left (1 + D a^2 \right) ~ ,
    \label{eq:Rud_fit}
 \ee
where the first term in the square brackets corresponds to the ratio of the SU(3) ChPT predictions at NNLO for the connected part of the light-quark contribution to $a_\ell^{\rm HVP, LO}$ in the infinite volume limit~\cite{Golowich:1995kd,Amoros:1999dp,Bijnens:2016ndo,Golterman:2017njs}, while the remaining terms parameterize the effects of the resummation of the higher orders, which, we remind, are not calculable within ChPT and are expected to dominate at the simulated pion masses (see Fig.~\ref{fig:ratioud_ETMC}).
The last term ($1 + D a^2$) takes into account possible discretization effects starting at order ${\cal{O}}(a^2)$ for our lattice setup. 

In Eq.~(\ref{eq:Rud_fit}) $A_0$, $A_1$ and $D$ are free parameters, while the ChPT terms contain two low-energy constants (LECs), $L_9^r$ and $C_{93}^r$.
We do not treat the latter ones as free parameters, but instead their values are fixed to the results of the analysis of $a_\mu^{\rm HVP, LO}(ud)$ performed in Ref.~\cite{Giusti:2018mdh}, namely $L_9^r(\mu = 0.77~\rm{GeV}) = 0.00273~(143)$ and $C_{93}^r(\mu = 0.77~\rm{GeV}) = -0.0136~(20)$ GeV$^{-2}$.
The uncertainties of the two LECs are properly taken into account through our bootstrap procedure (see Appendix~\ref{sec:appA}), which we remind has been adopted also in Ref.~\cite{Giusti:2018mdh}.

\begin{figure}[htb!]
\begin{center}
\includegraphics[scale=0.80]{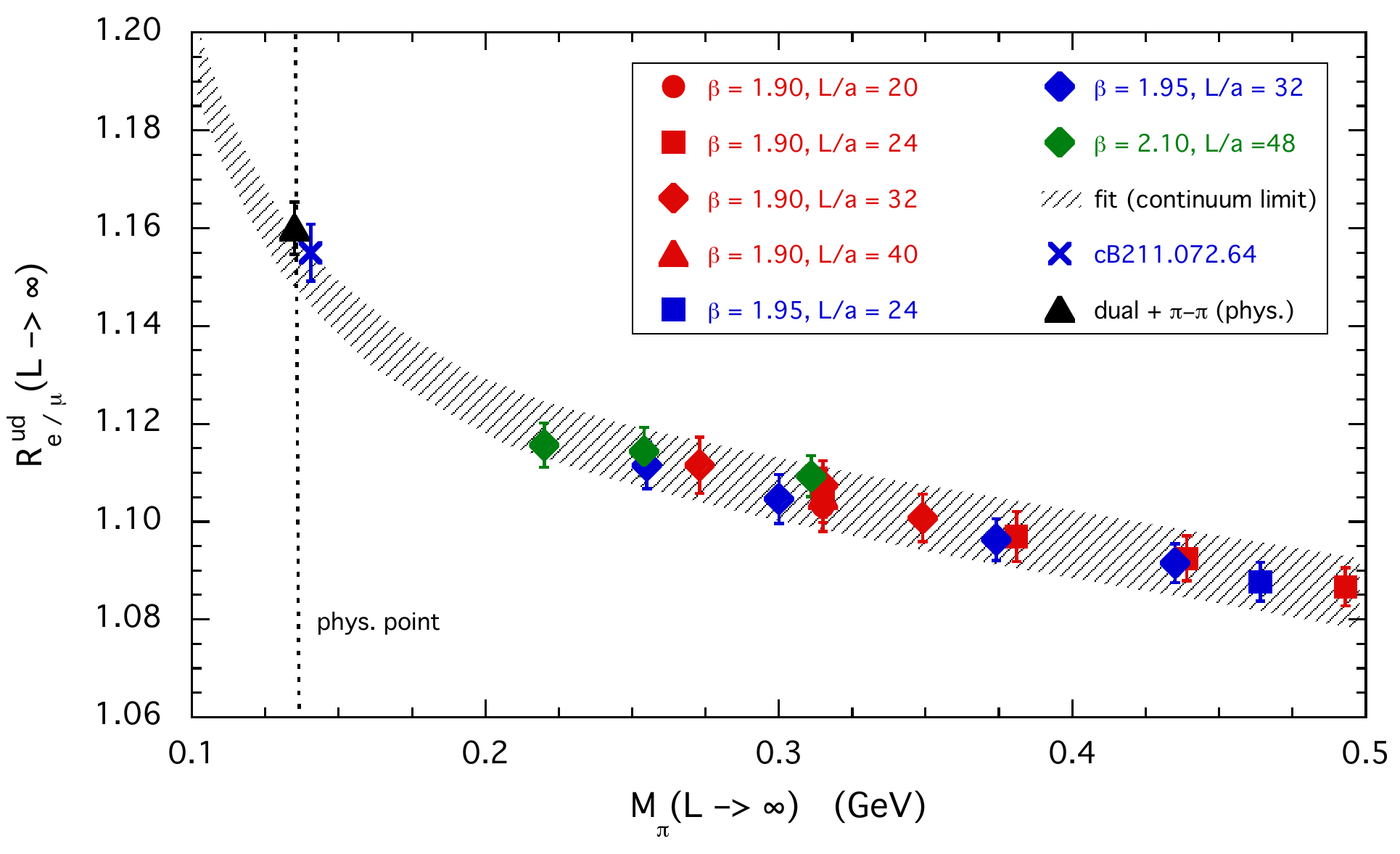}
\end{center}
\vspace{-1.0cm}
\caption{\it \small Values of the ratio $R_{e / \mu}^{ud}(L \to \infty)$ (see Eq.~(\ref{eq:Rud_FVE})) versus the simulated pion mass $M_\pi(L \to \infty)$ in the infinite volume limit for the $N_f = 2 + 1 + 1$ ETMC gauge ensembles of Appendix~\ref{sec:appA}. The shaded area corresponds to the results obtained with the fitting function (\ref{eq:Rud_fit}) at $1~\sigma$ level in the continuum limit. The blue cross represents the result corresponding to the new ETMC ensemble cB211.072.64 generated close to the physical pion mass~\cite{Alexandrou:2018egz} and corrected for FVEs (see text). The black triangle is the result obtained with the analytic representation of Ref.~\cite{Giusti:2018mdh} extrapolated to the physical pion mass and to the continuum and infinite volume limits (see Appendix~\ref{sec:appB}).
}
\label{fig:ratioud_ETMC_fit}
\end{figure}

The results obtained with the fitting function (\ref{eq:Rud_fit}), corresponding to a value of $\chi^2 / \mbox{d.o.f.} \simeq 0.2$ for $17$ data with $3$ free parameters, are shown in Fig.~\ref{fig:ratioud_ETMC_fit} by the shaded area, representing the fitting uncertainty at $1~\sigma$ level in the continuum limit.
Notice that discretization effects are almost negligible and overwhelmed by the uncertainties of the chiral extrapolation.
 Indeed, the value of the parameter $D$ in Eq.~(\ref{eq:Rud_fit}) is found to be compatible with zero.

At the physical pion mass and in the continuum limit the first strategy yields $R_{e / \mu}^{ud} = 1.1543~(54)$, where the error includes only the uncertainty induced by the statistical Monte Carlo errors of the simulations and its propagation in the fitting procedure. 
The above result shows that the chiral logs contained in the fitting function (\ref{eq:Rud_fit}) yield a significative enhancement of the ratio $R_{e / \mu}^{ud}$ toward the chiral limit.
We observe an effect of few percent, while the enhancement found in Ref.~\cite{Giusti:2018mdh} for $a_\mu^{\rm HVP, LO}(ud)$ due to the chiral extrapolation and to the continuum limit turned out to be much larger by almost an order of magnitude.

However, the chiral enhancement of $R_{e / \mu}^{ud}$ occurs in a region of pion masses not covered directly by the ETMC data.
Therefore, as our second strategy we make use of the recent ETMC ensemble, labelled cB211.072.64, generated with $N_f = 2 + 1 + 1$ dynamical quarks close to the physical pion mass ($M_\pi = 139\,(1)$ MeV) at a lattice spacing $a = 0.0803\,(4)$ fm and at a lattice size $L \simeq 5.1$ fm.
The lattice setup of the ensemble cB211.072.64 is described in details in Ref.~\cite{Alexandrou:2018egz} and briefly summarized in Appendix~\ref{sec:appA}.
The lattice action for cB211.072.64 differs from the one previously considered by ETMC in Refs.~\cite{Baron:2010bv,Baron:2011sf} by the addition of a Clover term.
Therefore, lattice artifacts may be different, but the presence of the Clover term turns out to be beneficial for reducing cutoff effects, in particular IB effects between the charged and the neutral pions~\cite{Alexandrou:2018egz}. 
We stress that discretization effects are expected to have a quite limited impact on the ratio $R_{e / \mu}^{ud}$.

According to Section II of Ref.~\cite{Giusti:2018mdh} and using 200 gauge configurations and 160 stochastic sources (diagonal in the spin variable and dense in the color one) per each gauge configuration, we have calculated the unrenormalized HVP terms $a_\ell^{\rm HVP, LO}(ud)$ for the electron and the muon, since the relevant renormalization constant (RC) of the local lattice version adopted for the em current operator is not yet available. 
Nevertheless, the electron to muon ratio does not depend on such RC, so that we get $R_{e / \mu}^{ud}({\rm cB211.072.64}) = 1.1414~(57)$.
After the subtraction of FVEs estimated through the analytic representation of $V^{ud}(t)$ evaluated at the physical pion mass, we get $R_{e / \mu}^{ud}({\rm cB211.072.64}, L \to \infty) = 1.1550~(58)$, which is shown in Fig.~\ref{fig:ratioud_ETMC_fit} as the blue cross and nicely confirms the chiral enhancement predicted by the fitting formula (\ref{eq:Rud_fit}).

As suggested by the smallness of the discretization effects exhibited by the data in Fig.~\ref{fig:ratioud_ETMC_fit}, it is interesting to use Eq.~(\ref{eq:Rud_fit}) for fitting the ETMC data obtained at the unphysical pion masses together with the result of the new cB211.072.64 ensemble close to the physical pion point without considering any discretization term, i.e.~by putting $D = 0$ in Eq.~(\ref{eq:Rud_fit}).
At the physical point, within the above second strategy, we get $R_{e / \mu}^{ud} = 1.1590~(56)$ with a $\chi^2 / \mbox{d.o.f.} \simeq 0.2$.

Finally, we adopt a third strategy based on the use of the analytic representation of the vector correlator $V^{ud}(t)$ developed in Ref.~\cite{Giusti:2018mdh}.
The main features of the representation are summarized in Appendix~\ref{sec:appB}. 
The crucial point is the extrapolation of the four parameters appearing in the representation (see Appendix~\ref{sec:appB} for their definitions) to the physical pion mass.
Correspondingly, we obtain the analytic representation of the vector correlator $V^{ud}(t)$ at the physical point.
We observe the following interesting facts.
\begin{itemize}

\item As shown in Ref.~\cite{Giusti:2018mdh}, the value of $a_\mu^{\rm HVP, LO}(ud)$ obtained after the chiral extrapolation of our analytic representation turned out to be consistent with the chiral extrapolation of the simulated values of $a_\mu^{\rm HVP, LO}(ud)$ inspired by ChPT.
This finding is remarkable and represents an evidence of the the reliability of the chiral extrapolation of our representation.

\item The first five moments of the polarization function have been evaluated using the $e^+ e^-$ data into two pions\footnote{Courtesy of A.~Keshavarzi, D.~Nomura and T.~Teubner.}. 
The corresponding predictions based on our analytic representation at the physical point were found to be nicely consistent with the above data~\cite{Giusti:2018mdh}. 
Also this finding is remarkable and represents a stringent test for the large time-distance tail of the vector correlator $V^{ud}(t)$, further reassuring about the reliability of the chiral extrapolation of our representation.

\item Our analytic representation includes not only the contribution of the isospin-1 $\pi - \pi$ spectrum, but also a dual part which nicely reproduces the vector correlator at short and intermediate time distances $t$. 
This means that our analytic representation provides the vector correlator $V^{ud}(t)$ for all values of $t$, not only for the discretized ones. 
This has the immediate consequence that the value of $a_\ell^{\rm HVP, LO}(ud)$ calculated by means of our representation at the physical point does not depend on the absolute scale setting.
This point is further elucidated at the end of the Appendix~\ref{sec:appB}.

\end{itemize}

Thus, besides the statistical uncertainties of the parameters appearing in the representation, the only important source of the systematic error comes from their chiral extrapolation.
For estimating the corresponding systematics we have tried several fitting functions and we have checked that it suffices to consider the four different fits in which: ~ i) the physical value of $M_\rho / M_\pi$ from PDG~\cite{PDG} is either included or not, and ~ ii) either a linear or a quadratic dependence on $m_{ud}$ is used to fit the dual energy $E_{dual}$.
The results obtained using the above four fitting choices are averaged according to Eq.~(28) of Ref.~\cite{Carrasco:2014cwa}.

Within the third strategy we get $R_{e / \mu}^{ud} = 1.1600 ~ (44) ~ (30) ~ [53]$, where the first error includes the uncertainties coming from the errors of the four parameters of the analytic representation and its propagation in the extrapolation to the physical point, the second error results from the different choices of their chiral extrapolation to the physical point, and the third error is their sum in quadrature.
The above result is shown in Fig.~\ref{fig:ratioud_ETMC_fit} as the black triangle.
Though being completely independent of the chiral enhancement found both for $a_\mu^{\rm HVP, LO}(ud)$ in Ref.~\cite{Giusti:2018mdh} and for the ratio $R_{e / \mu}^{ud}$, the result of the third strategy is nicely consistent both with the prediction of the fitting formula (\ref{eq:Rud_fit}) and with the result corresponding to the ensemble cB211.072.64.

By including in the systematic error the spread among the results of the three strategies (evaluated according to Eq.~(28) of Ref.~\cite{Carrasco:2014cwa}), our final determination of $R_{e / \mu}^{ud}$ is
\be
    R_{e / \mu}^{ud} = 1.1578 ~ (52)_{stat} ~ (39)_{syst} ~ [65] ~ . ~
    \label{eq:Remu_ud_final}
\ee

\subsection{Evaluation of $\widetilde{R}_{e / \mu}$}
\label{sec:Remu_other}

In this Section we determine the ratio $\widetilde{R}_{e / \mu}$, defined in Eq.~(\ref{eq:Remu_tilde}), corresponding to the LO HVP contributions other than the (connected) light-quark one.

We make use of a simple procedure based on the values of $a_\ell^{\rm HVP, LO}(ud)$, $a_\ell^{\rm HVP, LO}(s)$, $a_\ell^{\rm HVP, LO}(c)$ and $a_\ell^{\rm HVP, LO}(IB)$ obtained at the physical pion mass and in the continuum and infinite volume limits in Refs.~\cite{Giusti:2017jof,Giusti:2018mdh,Giusti:2019xct} and shown for $\ell = e, \mu$ in Tables 1 and 2 of Ref.~\cite{Giusti:2019hkz}.
In the case of the disconnected contribution $a_\ell^{\rm HVP, LO}(disc)$, following Refs.~\cite{Giusti:2018mdh,Giusti:2019hkz} we adopt the values $a_e^{\rm HVP, LO}(disc) = -3.80~(35) \cdot 10^{-14}$ from Ref.~\cite{Borsanyi:2017zdw} and $a_\mu^{\rm HVP, LO}(disc) = -12~(4) \cdot 10^{-10}$ from Refs.~\cite{Blum:2015you,Borsanyi:2017zdw,Blum:2018mom}.

The values of the ratios $a_\ell^{\rm HVP, LO}(j) / a_\ell^{\rm HVP, LO}(ud)$ for $j = s, c, IB, disc$ are collected in Table~\ref{tab:ratios} for $\ell = e, \mu$, where both the statistical and the systematic uncertainties are separately provided for each quantity.
In the case of the ratio $a_\ell^{\rm HVP, LO}(IB) / a_\ell^{\rm HVP, LO}(ud)$ the uncertainty includes also the estimate of quenching QED made in Ref.~\cite{Giusti:2019xct}.
We stress that the attractive features of the ratio $a_\ell^{\rm HVP, LO}(IB) / a_\ell^{\rm HVP, LO}(ud)$ are to be less sensitive to the uncertainties of the scale setting and to exhibit a reduced chiral dependence, which allows for a controlled, purely data-driven extrapolation to the physical point~\cite{Giusti:2019xct}.
\begin{table}[htb!]
\begin{center}
\begin{tabular}{||c|c|c|c|c||}
\hline
lepton & ~ $\frac{a_\ell^{\rm HVP, LO}(s)}{a_\ell^{\rm HVP, LO}(ud)} ~ $ & $ ~ \frac{a_\ell^{\rm HVP, LO}(c)}{a_\ell^{\rm HVP, LO}(ud)} ~ $ & $ ~ \frac{a_\ell^{\rm HVP, LO}(IB)}{a_\ell^{\rm HVP, LO}(ud)} ~ $ & $ ~ \frac{a_\ell^{\rm HVP, LO}(disc)}{a_\ell^{\rm HVP, LO}(ud)} ~ $ \\
\hline \hline
 e & ~ 0.0791~(34)~(36) ~ & ~ 0.0205~(10)~(8) ~ & ~ 0.0111~(28)~(51) ~ & ~ -0.0223~(15)~(14) ~ \\
\hline
 $\mu$ & ~ 0.0844~(30)~(32) ~ & ~ 0.0234~~(7)~(7) ~ & ~ 0.0113~(23)~(40) ~ & ~ -0.0191~(47)~(43) ~ \\ 
\hline   
\end{tabular}
\end{center}
\vspace{-0.25cm}
\caption{\it \small Values of the ratios $a_\ell^{\rm HVP, LO}(j) / a_\ell^{\rm HVP, LO}(ud)$ for $j = s, c, IB, disc$ obtained using the electron and muon results of Ref.~\cite{Giusti:2019hkz} (see text). For each entry the first and the second error represent the statistical and the systematic uncertainties, respectively.\hspace*{\fill}}
\label{tab:ratios}
\end{table}

We can now evaluate the ratio $\widetilde{R}_{e / \mu}$ by considering the four individual contributions corresponding to $j = s, c, IB, disc$ as $98 \%$ correlated between the numerator and the denominator.
The correlation is taken into account through a bootstrap procedure, which leads to the value
\be
   \widetilde{R}_{e / \mu} = 0.9895~(32)_{stat}~(31)_{syst}~[45] ~ ,
   \label{eq:Remu_other_final}
\ee
where the first error is statistical and the second one systematic, coming respectively from the separate statistical and systematic errors of the inputs of Table~\ref{tab:ratios}. 
The last error is their sum in quadrature.

New recent estimates of both $a_\mu^{\rm HVP, LO}(disc)$ and $a_\mu^{\rm HVP, LO}(IB)$ have been obtained in Ref.~\cite{Borsanyi:2020mff}.
The new determination $a_\mu^{\rm HVP, LO}(disc) = -15.4 ~(9) \cdot 10^{-10}$ is not inconsistent with the value $a_\mu^{\rm HVP, LO}(disc) = -12~(4) \cdot 10^{-10}$ we have adopted, while a significative reduction for $a_\mu^{\rm HVP, LO}(IB)$ is found. 
Such a difference may be due (at least partially) to the different prescriptions adopted to separate QCD and QED effects.
Nevertheless, even if we consider the case in which the IB contribution is completely dropped in the calculation, the value of $\widetilde{R}_{e / \mu}$ does not change significantly with respect to its uncertainty.

\section{Results}
\label{sec:results}

Collecting our findings (\ref{eq:Remu_ud_final}) and (\ref{eq:Remu_other_final}) our estimate of the electron-muon ratio $R_{e / \mu}$ is given by
\be
   R_{e / \mu} = \left( \frac{m_\mu}{m_e} \right)^2 ~ \frac{a_e^{\rm HVP, LO}}{a_\mu^{\rm HVP, LO}} = 1.1456 ~ (63)_{stat} ~ (54)_{syst} ~ [83] ~ ,
   \label{eq:Remu_final}
\ee
where the final error corresponds to a hadronic uncertainty of $\simeq 0.7 \%$, i.e.~a factor $\approx 4$ better than the individual precisions of the numerator and the denominator.

Our result (\ref{eq:Remu_final}) agrees very well with the one corresponding to the results $a_e^{\rm HVP, LO}(e^+ e^-) = 186.08~(0.66) \cdot 10^{-14}$ and $a_\mu^{\rm HVP, LO}(e^+ e^-) = 692.78~(2.42) \cdot 10^{-10}$ obtained from the dispersive analyses of $e^+ e^- \to$ hadrons data carried out recently in Ref.~\cite{Keshavarzi:2019abf}, namely
\be
   R_{e / \mu}^{e^+ e^-} = 1.1483 ~ (41)_e ~ (40)_\mu ~ [57] ~ ,
   \label{eq:Remu_disp}
\ee
where the first and second errors are related to the electron and muon contributions separately, while the third error is their sum in quadrature, i.e.~without taking into account correlations between the numerator and the denominator.
The uncertainty of the dispersive estimate of $R_{e / \mu}^{e^+ e^-}$ could be certainly reduced once the above correlations are properly taken into account.
For the purpose of the present work the conservative estimate of the error given in Eq.~(\ref{eq:Remu_disp}) is sufficient.

We can apply the procedure described in Section~\ref{sec:Remu_other} also to the individual results $a_\ell^{\rm HVP, LO}(j)$ for $j = ud, s, c, IB, disc$ obtained by the BMW Collaboration in Ref.~\cite{Borsanyi:2017zdw}. Assuming for sake of simplicity a $100 \%$ correlation between the individual contributions in the numerator and in the denominator we get 
\be
   R_{e / \mu} = 1.1381~(72) ~ ,
   \label{eq:Remu_BMW}
\ee
which is consistent within the uncertainties with our result (\ref{eq:Remu_final}) as well as with the dispersive one (\ref{eq:Remu_disp}).
Note that, as far as the individual term $a_\mu^{\rm HVP, LO}$ is concerned, the result of Ref.~\cite{Borsanyi:2017zdw} exhibits some tension with respect to both the ETMC result~\cite{Giusti:2018mdh,Giusti:2019xct,Giusti:2019hoy,Giusti:2019hkz} and the dispersive ones~\cite{Davier:2019can,Keshavarzi:2019abf,Jegerlehner:2017lbd}.
Moreover, the significance of such a tension is remarkably increased by the recent BMW result of Ref.~\cite{Borsanyi:2020mff}. 
However, the ratio $R_{e /\mu}$ is less sensitive to possible tensions between the results of various lattice collaborations and/or of dispersive analyses of $e^+ e^- \to$ hadrons data, which may occur for the individual hadronic terms $a_e^{\rm HVP, LO}$ and $a_\mu^{\rm HVP, LO}$.

\begin{figure}[htb!]
\begin{center}
\includegraphics[scale=0.75]{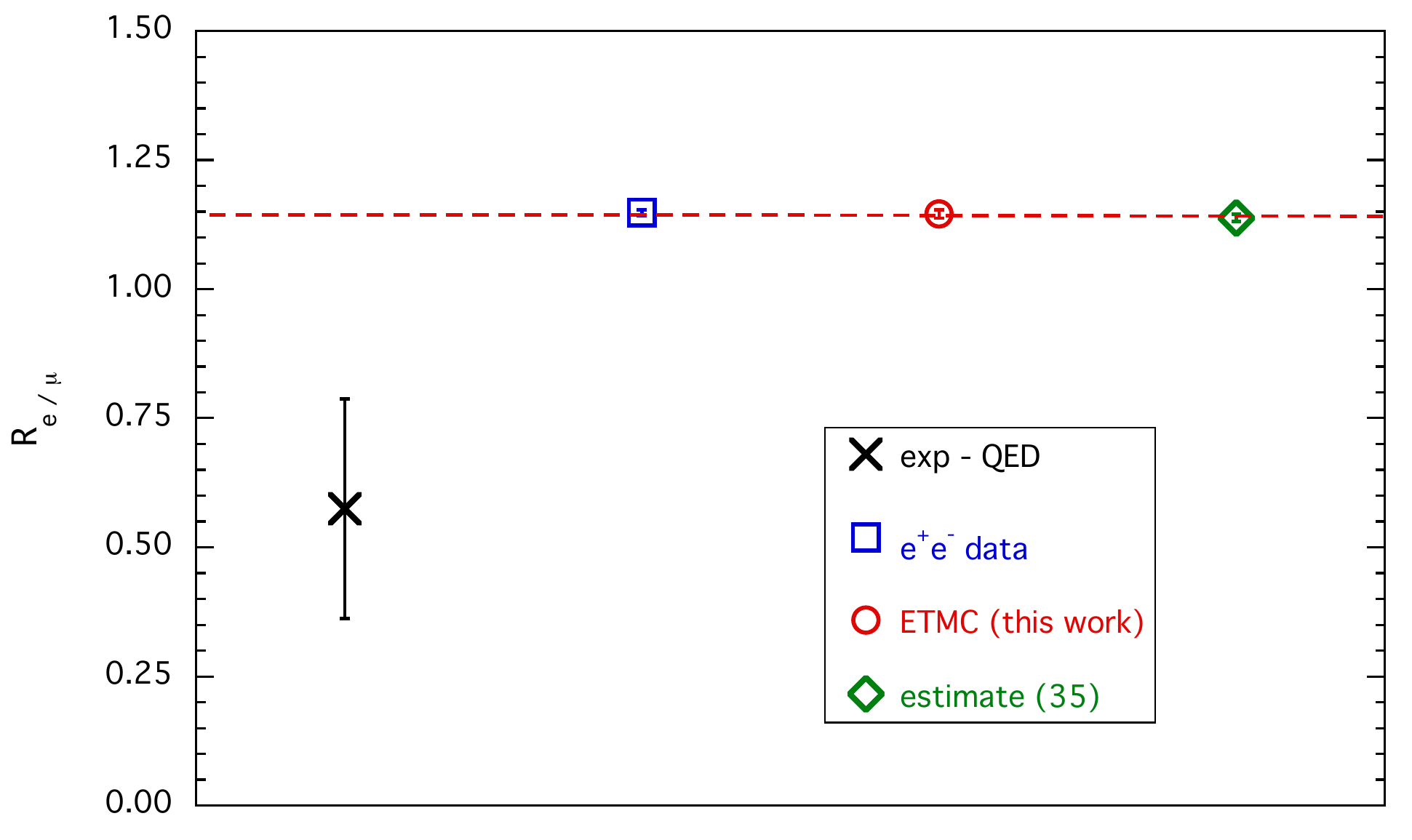}
\end{center}
\vspace{-1.0cm}
\caption{\it \small Comparison of the electron-muon ratio $R_{e/\mu}$  corresponding to the ``exp - QED'' estimate given by Eq.~(\ref{eq:ratio_mue_exp-QED}) (cross) with our lattice result (\ref{eq:Remu_final}) (red circle), the one given by Eq.~(\ref{eq:Remu_disp}) (blue square) derived from the results of the dispersive analyses of $e^+ e^- \to$ hadrons data carried out in Ref.~\cite{Keshavarzi:2019abf} and the estimate (\ref{eq:Remu_BMW}) (green diamond) obtained by applying the procedure of Section~\ref{sec:Remu_other} to the BMW results of Ref.~\cite{Borsanyi:2017zdw}. The dashed line corresponds to the central value of Eq.~(\ref{eq:Remu_final}).}
\label{fig:ratioemu}
\end{figure}

In Fig.~\ref{fig:ratioemu} our lattice result (\ref{eq:Remu_final}), the dispersive one (\ref{eq:Remu_disp}) and the estimate (\ref{eq:Remu_BMW}) are compared with the ``exp. - QED" value given by Eq.~(\ref{eq:ratio_mue_exp-QED}) in Section~\ref{sec:intro}.
As anticipated in Section~\ref{sec:intro} the ``exp. - QED" value differs from our lattice result (\ref{eq:Remu_final}) by $\simeq 2.7$ standard deviations.
An improvement by a factor of $\simeq 2$ in the precision of both the experiment and the QED contribution for the electron might be enough to reach a significance level of $\simeq 5$ standard deviations from our value (\ref{eq:Remu_final}).

Before closing this Section we provide our results also for the electron-$\tau$ and muon-$\tau$ ratios
 \be
    R_{e(\mu) / \tau} \equiv \left( \frac{m_\tau}{m_{e(\mu)}} \right)^2 ~ \frac{a_{e(\mu)}^{\rm HVP, LO}}{a_\tau^{\rm HVP, LO}} ~ .
    \label{eq:ratios_tau}
 \ee
We expect that the above ratios are more sensitive to the hadronic input $V(t)$, since the kinematical kernel $K_\ell(t)$ for the $\tau$-lepton differs significantly from the one of the electron(muon), as shown in Fig.~\ref{fig:kernels}.
Indeed the precision of our lattice data for the (connected) light-quark contributions $R_{\mu / \tau}(ud)$ and $R_{e / \tau}(ud)$ turns out to be at the level of $\approx 3 \%$, while the individual precisions for $a_{e,\mu, \tau}^{\rm HVP, LO}(ud)$ are at the level of $\approx 2 \%$ (see Ref.~\cite{Giusti:2019hkz}).
This result indicates that the numerator and the denominator in Eq.~(\ref{eq:ratios_tau}) can be considered almost uncorrelated.

The dependencies of $R_{e(\mu) / \tau}(ud)$ on the simulated pion mass, on the lattice spacing and on the lattice size is similar to the one shown in Fig.~\ref{fig:ratioud_ETMC} in the case of $R_{e / \mu}$.
The analyses of the data for both $R_{e(\mu) / \tau}(ud)$ and $\widetilde{R}_{e(\mu) / \tau}$ are similar to the ones described in Sections~\ref{sec:light} and \ref{sec:Remu_other} in the case of the corresponding electron-muon ratios, respectively.
The only difference is that the individual contributions corresponding to $j = ud, s, c, IB, disc$ should be considered to be uncorrelated between the numerator and the denominator.
In Table~\ref{tab:leptons} our final result for the three ratios $R_{e / \mu}$,  $R_{e / \tau}$ and $R_{\mu / \tau}$ are collected.
\begin{table}[htb!]
\begin{center}
\begin{tabular}{||c||c|c||}
\hline
~ $R_{e / \mu}$ ~ & ~ $R_{e / \tau}$ ~ & ~ $R_{\mu / \tau}$ ~ \\
\hline \hline
 ~ 1.1456~(83) ~ & ~ 6.69~(20) ~ & ~ 5.83~(17) ~ \\
\hline   
\end{tabular}
\end{center}
\vspace{-0.25cm}
\caption{\it \small Values of the ratios $R_{e / \mu}$,  $R_{e / \tau}$ and $R_{\mu / \tau}$ determined in this work at the physical pion mass and in the continuum and infinite volume limits. The errors include both statistical and systematic uncertainties added in quadrature.\hspace*{\fill}}
\label{tab:leptons}
\end{table}

Our findings for both $R_{e / \tau}$ and $R_{\mu / \tau}$ are consistent with the more precise ones corresponding to the recent results $a_e^{\rm HVP, LO}(e^+ e^-) = 186.08~(0.66) \cdot 10^{-14}$, $a_\mu^{\rm HVP, LO}(e^+ e^-) = 692.78~(2.42) \cdot 10^{-10}$ and $a_\tau^{\rm HVP, LO}(e^+ e^-) = 332.81~(1.39) \cdot 10^{-8}$ obtained from the dispersive analyses~\cite{Keshavarzi:2019abf} of $e^+ e^- \to$ hadrons data, namely
\bea
   \label{eq:Retau_disp}
   R_{e / \tau}^{e^+ e^-} & = & 6.760 ~ (24)_e ~ (28)_\tau ~ [37] ~ , ~ \\[2mm]
   \label{eq:Rmutau_disp}
   R_{\mu / \tau}^{e^+ e^-} & = & 5.887 ~ (21)_e ~ (25)_\tau ~ [33] ~ ,
\eea
where the third error is the sum in quadrature of the first two, i.e.~by considering the numerator and the denominator as uncorrelated.

\section{Conclusions}
\label{sec:conclusions}

In this work we have evaluated the ratios among the leading-order (LO) hadronic vacuum polarization (HVP) contributions to the anomalous magnetic moments of electron, muon and $\tau$-lepton, $a_{\ell = e, \mu, \tau}^{\rm HVP, LO}$, using lattice QCD+QED simulations. 
Our results include the effects at order ${\cal{O}}(\alpha_{em}^2)$ as well as the electromagnetic and strong-isospin breaking corrections at orders ${\cal{O}}(\alpha_{em}^3)$ and ${\cal{O}}(\alpha_{em}^2 (m_u - m_d))$, respectively, where $(m_u - m_d)$ is the $u$- and $d$-quark mass difference.
We have employed the gauge configurations generated by ETMC~\cite{Baron:2010bv,Baron:2011sf} with $N_f = 2+1+1$ dynamical quarks at three values of the lattice spacing ($a \simeq 0.062, 0.082, 0.089$ fm) with pion masses in the range $\simeq 210 - 450$ MeV.
The calculations are based on the quark-connected contributions to the HVP in the quenched-QED approximation, which neglects the charges of the sea quarks.
The quark-disconnected terms are estimated from results available in the literature~\cite{Borsanyi:2020mff,Borsanyi:2017zdw}.

We have shown that in the case of the electron-muon ratio the hadronic uncertainties in the numerator and in the denominator largely cancel out, while in the cases of the electron-$\tau$ and muon-$\tau$ ratios such a cancellation does not occur. 
At the physical pion mass and in the continuum and infinite volume limits we have obtained
 \bea
     \label{eq:Remu_ETMC}
     R_{e / \mu } & \equiv & \left( \frac{m_\mu}{m_e} \right)^2 ~ \frac{a_e^{\rm HVP, LO}}{a_\mu^{\rm HVP, LO}} = 1.1456~(83) ~ , \\[2mm]
     \label{eq:Retau_ETMC}
R_{e / \tau } & \equiv & \left( \frac{m_\tau}{m_e} \right)^2 ~ \frac{a_e^{\rm HVP, LO}}{a_\tau^{\rm HVP, LO}} = 6.69~(20) ~ , \\[2mm]
     \label{eq:Rmutau_ETMC}
R_{\mu / \tau } & \equiv & \left( \frac{m_\tau}{m_\mu} \right)^2 ~ \frac{a_\mu^{\rm HVP, LO}}{a_\tau^{\rm HVP, LO}} = 5.83~(17) ~ 
 \eea
with an uncertainty of $\simeq 0.7 \%$ for the electron-muon ratio and of $\simeq 3 \%$ for the electron-$\tau$ and muon-$\tau$ ratios.
Our results (\ref{eq:Remu_ETMC}-\ref{eq:Rmutau_ETMC}) agree very well with the corresponding estimates obtained using the recent results~\cite{Keshavarzi:2019abf} of the dispersive analyses of the experimental $e^+ e^- \to$ hadrons data (see Eqs.~(\ref{eq:Remu_disp}), (\ref{eq:Retau_disp}) and (\ref{eq:Rmutau_disp})).

We stress that the reduced sensitivity of $R_{e / \mu }$ to the hadronic uncertainties, present both in the numerator and in the denominator, makes our result (\ref{eq:Remu_ETMC}) an accurate SM prediction, weakening also possible tensions between the results of various lattice collaborations and/or of dispersive analyses of $e^+ e^- \to$ hadrons data.

Using the present determinations of the muon~\cite{Bennett:2006fi} and electron~\cite{Hanneke:2008tm,Hanneke:2010au} (g-2) experiments (see Eqs.~(\ref{eq:amu_exp}) and (\ref{eq:ae_exp})), the updated QED calculations~from Ref.~\cite{Aoyama:2019ryr} and the current estimates of the electro-weak, hadronic LBL and higher-order HVP contributions, the ``exp - QED'' value of the electron-muon ratio $R_{e / \mu}$ (see Eq.~(\ref{eq:ratio_mue_exp-QED}) of Section~\ref{sec:intro}) is equal to
\be
     R_{e / \mu}^{\rm exp - QED} = 0.575 ~ (213)_e ~ (6)_\mu ~ [213] ~ ,
     \label{eq:ratioemu_exp-QED}
\ee
which differs from our SM result (\ref{eq:Remu_ETMC}) by $\simeq 2.7$ standard deviations.
We stress that such a tension is dominated by present experimental and QED uncertainties, while the role of the hadronic uncertainties on $R_{e / \mu }$ is quite marginal.
Thus, an improvement by a factor of $\simeq 2$ in the precision of both the experiment and the QED contribution to the electron ($g - 2$) could be enough to reach a tension with the SM at a significance level of $\simeq 5$ standard deviations.

\section*{Acknowledgments}

We gratefully acknowledge C.~Lehner and M.~Hoferichter for useful comments and V.~Lubicz for a careful reading of the manuscript.
We warmly thank F.~Sanfilippo for providing us the code for calculating the light-quark contribution to the vector current-current correlator in the case of the ETMC gauge ensemble cB211.072.64~\cite{Alexandrou:2018egz}.
We thank also B.~Kostrezwa for his help in the production of the gauge ensemble A40.40 with the tmLQCD software package~\cite{Jansen:2009xp,Abdel-Rehim:2013wba,Deuzeman:2013xaa}.

\appendix

\section{Lattice framework and simulation details}
\label{sec:appA}

The gauge ensembles used in this work are those generated by ETMC with $N_f = 2 + 1 + 1$ dynamical quarks~\cite{Baron:2010bv,Baron:2011sf} and used in Ref.~\cite{Carrasco:2014cwa} to determine the up, down, strange and charm quark masses. 
We use the Iwasaki action~\cite{Iwasaki:1985we} for the gluons and the Wilson Twisted Mass Action~\cite{Frezzotti:2000nk,Frezzotti:2003xj,Frezzotti:2003ni} for the sea quarks. 
In the valence sector we adopt a non-unitary setup~\cite{Frezzotti:2004wz} in which the strange quark is regularized as an Osterwalder-Seiler fermion~\cite{Osterwalder:1977pc}, while the up and down quarks have the same action as the sea.
Working at maximal twist such a setup guarantees an automatic ${\cal{O}}(a)$-improvement~\cite{Frezzotti:2003ni,Frezzotti:2004wz}.

We have performed simulations at three values of the inverse bare lattice coupling $\beta$ and at several different lattice volumes as shown in Table~\ref{tab:simudetails}. 
We allow a separation of 20 trajectories between each of the $N_{\mathrm{cfg}}$ analysed configurations.
For the earlier investigation of finite-volume effects (FVEs) ETMC had produced three dedicated ensembles, A40.20, A40.24 and A40.32, which share the same quark masses and lattice spacing and differ only in the lattice size $L$.
To improve such an investigation a further gauge ensemble, A40.40, has been produced at a larger value of the lattice size $L$.

\begin{table}[htb!]
{\scriptsize
\begin{center}
\begin{tabular}{||c|c|c|c||c|c|c||c|c||c|c|c||c||}
\hline
ensemble & $\beta$ & $V / a^4$ &$N_{\mathrm{cfg}}$&$a\mu_{sea}=a\mu_{ud}$&$a\mu_\sigma$&$a\mu_\delta$& $a\mu_s$ & $a\mu_c$ & $M_\pi {\rm (MeV)}$ & $M_K {\rm (MeV)}$ & $M_D {\rm (MeV)}$ & $M_\pi L$ \\
\hline \hline
$A40.40$ & $1.90$ & $40^{3}\times 80$ &$100$ &$0.0040$ &$0.15$ &$0.19$ & $0.02363$ & $0.27903$ & 317 (12) & 576 (22) & 2002 (77) & 5.7 \\
\cline{1-1} \cline{3-5} \cline{10-13}
$A30.32$ & & $32^{3}\times 64$ &$150$ &$0.0030$ & & & & & 275 (10) & 568 (22) & 2012 (77) & 3.9 \\
$A40.32$ & & & $100$ & $0.0040$ & & & & & 316 (12) & 578 (22) & 2008 (77) & 4.5 \\
$A50.32$ & & & $150$ & $0.0050$ & & & & & 350 (13) & 586 (22) & 2014 (77) & 5.0 \\
\cline{1-1} \cline{3-5} \cline{10-13}
$A40.24$ & & $24^{3}\times 48 $ & $150$ & $0.0040$ & & & & & 322 (13) & 582 (23) & 2017 (77) & 3.5 \\
$A60.24$ & & & $150$ & $0.0060$ & & & & & 386 (15) & 599 (23) & 2018 (77) & 4.2 \\
$A80.24$ & & & $150$ & $0.0080$ & & & & & 442 (17) & 618 (14) & 2032 (78) & 4.8 \\
$A100.24$ & & & $150$ & $0.0100$ & & & & & 495 (19) & 639 (24) & 2044 (78) & 5.3 \\
\cline{1-1} \cline{3-5} \cline{10-13}
$A40.20$ & & $20^{3}\times 48 $ & $150$ & $0.0040$ & & & & & 330 (13) & 586 (23) & 2029 (79) & 3.0 \\
\hline \hline
$B25.32$ & $1.95$ & $32^{3}\times 64$ & $150$ &$0.0025$&$0.135$ &$0.170$ & $0.02094$ & $0.24725$ & 259 ~(9) & 546 (19) & 1942 (67) & 3.4 \\
$B35.32$ & & & $150$ & $0.0035$ & & & & & 302 (10) & 555 (19) & 1945 (67) & 4.0 \\
$B55.32$ & & & $150$ & $0.0055$ & & & & & 375 (13) & 578 (20) & 1957 (68) & 5.0 \\
$B75.32$ & & & $~80$ & $0.0075$ & & & & & 436 (15) & 599 (21) & 1970 (68) & 5.8 \\
\cline{1-1}\cline{3-5} \cline{10-13}
$B85.24$ & & $24^{3}\times 48 $ & $150$ & $0.0085$ & & & & & 468 (16) & 613 (21) & 1972 (68) & 4.6 \\
\hline \hline
$D15.48$ & $2.10$ & $48^{3}\times 96$ & $100$ & $0.0015$ & $0.1200$ & $0.1385$ & $0.01612$ & $0.19037$ & 223 ~(6) & 529 (14) & 1929 (49) & 3.4 \\ 
$D20.48$ & & & $100$ & $0.0020$ & & & & & 256 ~(7) & 535 (14) & 1933 (50) & 3.9 \\
$D30.48$ & & & $100$ & $0.0030$ & & & & & 312 ~(8) & 550 (14) & 1937 (49) & 4.7 \\
 \hline   
\end{tabular}
\end{center}
}
\vspace{-0.25cm}
\caption{\it \small Values of the valence and sea bare quark masses (in lattice units), of the pion, kaon and D-meson masses for the $N_f = 2+1+1$ ETMC gauge ensembles used in Ref.~\cite{Carrasco:2014cwa} and for the gauge ensemble, A40.40 added to improve the investigation of FVEs. A separation of $20$ trajectories between each of the $N_{\mathrm{cfg}}$ analysed configurations. The bare twisted masses $\mu_\sigma$ and $\mu_\delta$ describe the strange and charm sea doublet as in to Ref.~\cite{Frezzotti:2003xj}. The values of the strange and charm quark bare masses $a \mu_s$ and $a \mu_c$, given for each $\beta$, correspond to the physical strange and charm quark masses, $m_s^{\textrm{phys}}(\overline{\rm MS}, 2\,\mbox{{\rm GeV}}) = 99.6 (4.3)$ MeV and $m_c^{\textrm{phys}}(\overline{\rm MS}, 2\,\mbox{{\rm GeV}}) = 1176 (39)$ MeV, and to the mass RCs determined in Ref.~\cite{Carrasco:2014cwa}. The central values and errors of pion, kaon and D-meson masses are evaluated using the bootstrap procedure of Ref.~\cite{Carrasco:2014cwa}. The two valence quarks in the pseudoscalar mesons are regularized with opposite values of the Wilson $r$-parameter in order to guarantee that discretisation effects on the meson masses are of order ${\cal{O}}(a^2 \mu ~ \Lambda_{\textrm{QCD}})$.\hspace*{\fill}}
\label{tab:simudetails}
\end{table}

At each lattice spacing, different values of the light sea-quark masses have been considered. 
The light valence and sea quark masses are always taken to be degenerate. 
The bare masses of the valence strange and charm quarks ($a\mu_s$ and $a\mu_c$) are obtained, at each $\beta$, using the physical strange and charm masses and the mass RCs determined in Ref.~\cite{Carrasco:2014cwa}. 
There the ``FLAG" hadronic scheme was adopted in which the pion and kaon masses in isosymmetric QCD are equal to $M_\pi^{(0),\textrm{FLAG}} = 134.98$ MeV and $M_K^{(0),\textrm{FLAG}} = 494.2~(4)$ MeV and the lattice scale is fixed by the value $f_\pi^{(0),\textrm{FLAG}} = 130.41~(20)$ MeV for the physical pion decay constant. 
In the charm sector instead, the $D_s$-meson mass $M_{D_s}^{(0)}$ was chosen to be equal to its experimental value $M_{D_s^+} = 1969.0(1.4)$ MeV~\cite{PDG}.
The values of the lattice spacing are found to be: $a = 0.0885(36)$, $0.0815(30)$, $0.0619(18)$ fm at $\beta = 1.90$, $1.95$ and $2.10$, respectively.
In Ref.~\cite{Giusti:2017dwk} it was shown that at the current level of precision the ``FLAG" hadronic scheme is equivalent to the Gasser-Rusetsky-Scimemi prescription~\cite{Gasser:2003hk}.

In this work, as well as in all our works on the muon HVP terms~\cite{Giusti:2017jof,Giusti:2018mdh,Giusti:2019xct}, we made use of the bootstrap samples generated for the input parameters of the quark mass analysis of Ref.~\cite{Carrasco:2014cwa}.
There, eight branches of the analysis were adopted differing in: 
\begin{itemize}
\item the continuum extrapolation adopting for the matching of the lattice scale either the Sommer parameter $r_0$ or the mass of a fictitious P-meson made up of two valence strange(charm)-like quarks; 
\item the chiral extrapolation performed with fitting functions chosen to be either a polynomial expansion or a Chiral Perturbation Theory (ChPT) Ansatz in the light-quark mass;
\item the choice between the methods M1 and M2, which differ by ${\cal{O}}(a^2)$ effects, used to determine the mass RC $Z_m = 1 / Z_P$ in the RI$^\prime$-MOM scheme. 
\end{itemize}

Statistical errors on the meson masses and the various HVP terms are evaluated  using the jackknife procedure. 
The uncertainties based on data obtained from independent ensembles of gauge configurations, like the errors of the fitting procedures, are evaluated using the above bootstrap events 
in order to take properly into account cross-correlations.
The results corresponding to the eight branches of the analysis are then averaged according to Eq.~(28) of Ref.~\cite{Carrasco:2014cwa}.

The statistical accuracy of the meson correlator is based on the use of the so-called ``one-end'' stochastic method~\cite{McNeile:2006bz}, which includes spatial stochastic sources at a single time slice chosen randomly.
In the case of the light-quark contribution we have used 160 stochastic sources (diagonal in the spin variable and dense in the color one) per each gauge configuration, while for the strange (charm) quark contribution 4(1) stochastic sources have been employed per each gauge configuration.

In Table~\ref{tab:data_ud} we have collected the results for the (connected) light-quark contribution to the electron-muon ratio $R_{e / \mu}^{ud}$ (see Eq.~(\ref{eq:Remu_ud})), both at finite volume and in infinite volume limit, $R_{e / \mu}^{ud}(L \to \infty)$, evaluated according to Eq.~(\ref{eq:Rud_FVE}) using the procedure of Ref.~\cite{Giusti:2018mdh} for removing FVEs on both the pion masses and the lepton HVP terms (see Section~\ref{sec:light}) for each of the ETMC gauge ensembles of Table~\ref{tab:simudetails}. 
\begin{table}[htb!]
{\small
\begin{center}
\begin{tabular}{||c|c|c|c||c|c||c|c||}
\hline
ensemble & $\beta$ & $V / a^4$ &$a\mu_{sea}=a\mu_{ud}$& ~~ $M_\pi(L)$ ~~ & ~~ $R_{e / \mu}^{ud}(L)$ ~~ & $M_\pi(L \to \infty)$ & $R_{e / \mu}^{ud}(L \to \infty)$ \\
\hline \hline
$A40.40$ & $1.90$ & $40^{3}\times 80$ & $0.0040$ & 317 (12) & 1.1067 (54) &  315 (13) & 1.1071 (53) \\
\cline{1-1} \cline{3-8}
$A30.32$ & & $32^{3}\times 64$ & $0.0030$ & 275 (10) & 1.1075 (65) & 273 (10) & 1.1115 (57) \\
$A40.32$ & & &  $0.0040$ & 316 (12) & 1.1056 (57) & 315 (13) & 1.1074 (51) \\
$A50.32$ & & & $0.0050$ & 350 (13) & 1.0971 (45) & 349 (13) & 1.1008 (49) \\
\cline{1-1} \cline{3-8}
$A40.24$ & & $24^{3}\times 48 $ & $0.0040$ & 322 (13) & 1.0980 (66) & 315 (13) & 1.1059 (49) \\
$A60.24$ & & & $0.0060$ & 386 (15) & 1.0898 (51) & 381 (15) & 1.0970 (51) \\
$A80.24$ & & & $0.0080$ & 442 (17) & 1.0887 (48) & 439 (17) & 1.0925 (46)  \\
$A100.24$ & & & $0.0100$ & 495 (19) & 1.0831 (44) & 493 (19) & 1.0867 (39) \\
\cline{1-1} \cline{3-8} 
$A40.20$ & & $20^{3}\times 48 $ & $0.0040$ & 330 (13) & 1.0886 (61) & 315 (13) & 1.1031 (51) \\
\hline \hline
$B25.32$ & $1.95$ & $32^{3}\times 64$ &$0.0025$& 259 ~(9) & 1.1053 (56) & 255 ~(9) & 1.1115 (47) \\
$B35.32$ & & & $0.0035$ & 302 (10) & 1.1029 (51) & 300 (10) & 1.1046 (50) \\
$B55.32$ & & & $0.0055$ & 375 (13) & 1.0947 (44) & 374 (13) & 1.0963 (43) \\
$B75.32$ & & & $0.0075$ & 436 (15) & 1.0912 (37) & 435 (15) & 1.0915 (39) \\
\cline{1-1}\cline{3-8}
$B85.24$ & & $24^{3}\times 48 $ & $0.0085$ & 468 (16) & 1.0870 (38) & 464 (16) & 1.0877 (40) \\
\hline \hline
$D15.48$ & $2.10$ & $48^{3}\times 96$ & $0.0015$ & 223 ~(6) & 1.1108 (66) & 220 ~(6) & 1.1156 (45) \\ 
$D20.48$ & & & $0.0020$ & 256 ~(7) & 1.1067 (68) & 254 ~(7) & 1.1143 (50) \\
$D30.48$ & & & $0.0030$ & 312 ~(8) & 1.1073 (46) & 311 ~(8) & 1.1093 (42) \\
 \hline   
\end{tabular}
\end{center}
}
\vspace{-0.25cm}
\caption{\it \small Values of the (connected) light-quark contribution to the electron-muon ratio  (see Eq.~(\ref{eq:Remu_ud})) both at finite volume, $R_{e / \mu}^{ud}(L)$, and in the infinite volume limit, $R_{e / \mu}^{ud}(L \to \infty)$, evaluated according to Eq.~(\ref{eq:Rud_FVE}) using the procedure of Ref.~\cite{Giusti:2018mdh} for removing FVEs (see Section~\ref{sec:light}) for each of the ETMC gauge ensembles of Table~\ref{tab:simudetails}. Pion masses both at finite volume and in the infinite volume limit, evaluated according to Ref.~\cite{Giusti:2018mdh}, are given in MeV. All the errors include (in quadrature) both the statistical and the systematic uncertainties corresponding to the bootstrap samples of Ref.~\cite{Carrasco:2014cwa}.\hspace*{\fill}}
\label{tab:data_ud}
\end{table}
We stress that, thanks to our bootstrap samples, the uncertainties on the pion masses are properly propagated in our fitting procedure of Section~\ref{sec:light}.

For the second strategy adopted in Section~\ref{sec:light} to test the chiral extrapolation of $R_{e / \mu}^{ud}$ we have used 200 gauge configurations of the ensemble $cB211.072.64$ generated by ETMC with $N_f = 2+1+1$ dynamical quarks close to the physical pion mass~\cite{Alexandrou:2018egz}.
The gauge action is still the Iwasaki action~\cite{Iwasaki:1985we}, but the fermionic (twisted-mass) actions in both light and heavy sectors contain an additional Clover term with a Sheikoleslami-Wohlert~\cite{Sheikholeslami:1985ij} improvement coefficient $c_{SW}$ taken from 1-loop tadpole boosted perturbation theory~\cite{Aoki:1998qd}.
The presence of the Clover term turns out to be beneficial for reducing cutoff effects, in particular IB effects between the charged and the neutral pions.
The masses of the two degenerate light quarks, of the strange and charm quarks are tuned to their physical values.
The simulated pion mass turns out to be equal to $M_\pi = 139\,(1)$ MeV and the lattice spacing is estimated to be $a = 0.0803\,(4)$ fm using as input both mesonic and baryonic quantities.
The lattice volume is $V = 64^3 \times 128 ~ a^4$, so that the product $M_\pi L$ is equal to $\simeq 3.6$.

\section{The dual + $\pi \pi$ representation of the vector correlator $V^{ud}(t)$}
\label{sec:appB}

Following Ref.~\cite{Giusti:2018mdh} the analytic representation, $V_{dual + \pi \pi}(t)$, of the (connected) light-quark contribution $V^{ud}(t)$ is given by the sum of two terms
\be
      V_{dual + \pi \pi}(t) \equiv V_{dual}(t) + V_{\pi \pi}(t) ~ ,
      \label{eq:dual+2pi}
 \ee
where $V_{\pi \pi}(t)$ represents the two-pion contribution in a finite box, while $V_{dual}(t)$ is the ``dual'' representation of the tower of the contributions coming from the excited states above the two-pion ones.
Therefore, $V_{\pi \pi}(t)$ is expected to dominate at large time distances $t$, while the contribution of $V_{dual}(t)$ is crucial at low and intermediate time distances, as firstly observed in Ref.~\cite{Giusti:2017jof}.

The correlator $V_{dual}(t)$ is defined as
\be
    V_{dual}(t) \equiv \frac{1}{24 \pi^2} R_{dual} \int_{s_{dual}}^\infty ds \sqrt{s} e^{-\sqrt{s} t} R^{pQCD}(s) ~ ,
     \label{eq:Vdual_0}
\ee
where $s_{dual}$ is an effective threshold {\it \`a la SVZ}, above which the hadronic spectral density is dual to the perturbative QCD (pQCD) prediction $R^{pQCD}(s)$ of the $e^+ e^-$ cross section into hadrons, while $R_{dual}$ is a multiplicative factor introduced mainly to take into account discretization effects.
According to the traditional QCD sum rule framework~\cite{SVZ} the value of $\sqrt{s_{dual}}$ is expected to be above the ground-state mass of the relevant channel by an amount of the order of $\Lambda_{QCD}$.
Therefore, following Ref.~\cite{Giusti:2018mdh} we assume that $s_{dual} = \left( M_\rho + E_{dual} \right)^2$ with $M_\rho$ being the mass of the $\rho$-meson vector resonance and $E_{dual}$ a parameter of order $\Lambda_{QCD}$. 

Since the effective threshold $s_{dual}$ is well above the light-quark threshold $4 m_{ud}^2$, the pQCD density $R^{pQCD}(s)$  is dominated by its leading term of order ${\cal{O}}(\alpha_s^0)$ in the relevant range of the integration over $s$ in the r.h.s.~of Eq.~(\ref{eq:Vdual_0}).
Higher-order corrections (as well as condensates and the slight dependence on the light-quark mass $m_{ud}$) should play a sub-leading role and they can be taken into account by the effective parameter $R_{dual}$ in Eq.~(\ref{eq:Vdual_0}).

The dual correlator $V_{dual}(t)$ can be explicitly written as~\cite{Giusti:2018mdh}
\be
      V_{dual}(t) = \frac{5}{18 \pi^2} \frac{R_{dual}}{t^3} e^{- (M_\rho + E_{dual}) t} \left[ 1 + (M_\rho + E_{dual}) t +
                           \frac{1}{2} (M_\rho + E_{dual})^2 t^2 \right] ~ ,
      \label{eq:Vdual}
 \ee
where $R_{dual}$, $E_{dual}$ and $M_\rho$ are free parameters to be determined by fitting the lattice data for the light-quark vector correlator $V^{ud}(t)$. 
Note that the $\rho$-meson mass $M_\rho$ will appear also in the two-pion contribution $V_{\pi \pi}(t)$.

As it is well known after Refs.~\cite{Luscher:1985dn,Luscher:1986pf,Luscher:1990ux,Luscher:1991cf}, the energy levels $\omega_n$ of two pions in a finite box of volume $L^3$ are given by
 \be
     \omega_n = 2 \sqrt{M_\pi^2 + k_n^2} ~ ,
     \label{eq:omegan}
 \ee
where the discretized values $k_n$ should satisfy the L\"uscher condition, which for the case at hand (two pions in a $P$-wave with total isospin $1$) reads as 
 \be
     \delta_{11}(k_n) + \phi\left( \frac{k_nL}{2\pi} \right) = n \pi ~ ,
     \label{eq:kn}
 \ee
with $\delta_{11}$ being the (infinite volume) scattering phase shift and $\phi(z)$ a known kinematical function given by
\be
    \mbox{tan}\phi(z) = - \frac{2 \pi^2 z}{\sum_{\vec{m} \in \mathbb{Z}^3} \left( |\vec{m}|^2 - z^2 \right)^{-1}} ~ .
    \label{eq:phi}
\ee

The two-pion contribution $V_{\pi \pi}(t)$ can be written as~\cite{Lellouch:2000pv,Meyer:2011um,Francis:2013qna}
 \be
     V_{\pi \pi}(t) = \sum_n \nu_n |A_n|^2 e^{-\omega_n t} ~ ,
     \label{eq:V2pi}
 \ee
where $\nu_n$ is the number of vectors $\vec{z} \in \mathbb{Z}^3$ with norm $|\vec{z}|^2 = n$ and the squared amplitudes $|A_n|^2$ are related to the timelike pion form factor $F_\pi(\omega) = |F_\pi(\omega_n)| e^{i\delta_{11}(k_n)}$ by
 \be
     \nu_n |A_n|^2 = \frac{2 k_n^5}{3 \pi \omega_n^2} |F_\pi(\omega_n)|^2\left[ k_n \delta_{11}^\prime(k_n) + 
                               \frac{k_nL}{2\pi} \phi^\prime\left( \frac{k_nL}{2\pi} \right) \right]^{-1} ~ .
     \label{eq:An}
 \ee
Following Ref.~\cite{Giusti:2018mdh} we adopt the Gounaris-Sakurai (GS) parameterization~\cite{Gounaris:1968mw}, which is based on the dominance of the $\rho$-meson resonance in the amplitude of the pion-pion P-wave elastic scattering (with total isospin $1$), namely
\be
     F_\pi^{(GS)}(\omega) = \frac{M_\rho^2 - A_{\pi \pi}(0)}{M_\rho^2 - \omega^2 - A_{\pi \pi}(\omega)} ~ ,
     \label{eq:Fpi_GS}
\ee
where the (twice-subtracted~\cite{Gounaris:1968mw}) pion-pion amplitude $A_{\pi \pi}(\omega)$ is given by
\be
    A_{\pi \pi}(\omega) = h(M_\rho) + (\omega^2 - M_\rho^2) \frac{h^\prime(M_\rho)}{2 M_\rho} - h(\omega) + i \omega \Gamma_{\rho \pi \pi}(\omega)
    \label{eq:Apipi}
\ee
with
\bea
    \label{eq:Gamma_rhopipi}
    \Gamma_{\rho \pi \pi}(\omega) & = & \frac{g_{\rho \pi \pi}^2}{6 \pi} \frac{k^3}{\omega^2} ~ , \\[2mm]
    \label{eq:homega}
    h(\omega) & = & \frac{g_{\rho \pi \pi}^2}{6 \pi} \frac{k^3}{\omega} \frac{2}{\pi}\mbox{log}\left( \frac{\omega + 2k}{2M_\pi} \right) ~ , \\[2mm]
    \label{eq:hpomega}
    h^\prime(\omega) & = & \frac{g_{\rho \pi \pi}^2}{6 \pi} \frac{k^2}{\pi \omega} \left\{ 1 + \left(1 + \frac{2 M_\pi^2}{\omega^2} \right) 
                                           \frac{\omega}{k} \mbox{log}\left(\frac{\omega + 2k}{2 M_\pi} \right) \right\} ~ , \\[2mm]
    \label{eq:Apipi0}
     A_{\pi \pi}(0) & = & h(M_\rho) - \frac{M_\rho}{2} h^\prime(M_\rho) + \frac{g_{\rho \pi \pi}^2}{6 \pi} \frac{M_\pi^2}{\pi}                                   
\eea
and $k \equiv \sqrt{\omega^2 / 4 - M_\pi^2}$.
By analytic continuation the GS form factor at $\omega = 0$ is normalized to unity, i.e.~$F_\pi^{(GS)}(\omega = 0) = 1$.
The scattering phase shift $\delta_{11}(k)$, i.e.~the phase of the pion form factor according to the Watson theorem, is given by
\be
    \mbox{cot}\delta_{11}(k) = \frac{M_\rho^2 - \omega^2 - h(M_\rho) - (\omega^2 - M_\rho^2) h^\prime(M_\rho) / (2 M_\rho) +
                                              h(\omega)}{\omega \Gamma_{\rho \pi \pi}(\omega)} ~ .
    \label{eq:delta11}
\ee

The GS form factor (\ref{eq:Fpi_GS}) contains two parameters: the resonance mass $M_\rho$ and its strong coupling with two pions $g_{\rho \pi \pi}$.
Together with $R_{dual}$ and $E_{dual}$, appearing in the dual contribution (\ref{eq:Vdual}), they have been determined in Ref.~\cite{Giusti:2018mdh} by fitting the lattice data for the light-quark correlator $V^{ud}(t)$ for each of the ETMC ensembles of Appendix~\ref{sec:appA}.

More precisely, for each lattice spacing and volume the following dimensionless parameters, $R_{dual}$, $E_{dual} / M_\pi$, $M_\rho / M_\pi$ and $g_{\rho \pi \pi}$, are determined by fitting the data for $V^{ud}(t)$ in lattice units and the knowledge of the value of the lattice spacing is not required.
In this way all the four parameters of the analytic representation~(\ref{eq:dual+2pi}) were determined as a function of the light-quark mass $m_{ud}$, lattice spacing $a$ and lattice size $L$ together with their statistical+fitting uncertainties.
As shown in Ref.~\cite{Giusti:2018mdh} the above dependencies, in particular the one related to the light-quark mass $m_{ud}$, are much less problematic for the parameters of the representation~(\ref{eq:dual+2pi}) with respect to the quantity $a_\mu^{\rm HVP, LO}(ud)$ itself.
Moreover, the infinite volume limit was performed at each simulated light-quark mass $m_{ud}$ and lattice spacing $a$, obtaining in this way a proper evaluation of FVEs. 
The four parameters were extrapolated to the physical pion mass and to the continuum and infinite volume limits, namely $R_{dual}^{phys}$, $(E_{dual} / M_\pi)^{phys}$, $(M_\rho / M_\pi)^{phys}$ and $g_{\rho \pi \pi}^{phys}$.
These values do not depend on the absolute scale setting, but only on the relative ones.
Correspondingly, the analytic representation of the vector correlator was obtained at the physical point, $V_{dual + \pi \pi}^{phys}(t)$.

We want to highlight an important feature of the representation $V_{dual + \pi \pi}^{phys}(t)$, related to the fact that it is determined for all values of the time distance $t$, not only for the discretized ones.
This has the immediate consequence that the value of $a_\ell^{\rm HVP, LO}(ud)$ calculated by means of $V_{dual + \pi \pi}^{phys}(t)$ does not depend on the absolute scale setting.
As well known (see, e.g., Ref.~\cite{Aoyama:2020ynm}), this may represent an important source of uncertainty in the evaluation of $a_\ell^{\rm HVP, LO}(ud)$ using the discretized lattice data for the vector correlator.
The main observation is that the analytic representation $V_{dual + \pi \pi}(t)$ can be written as 
\be
    V_{dual + \pi \pi}(t) = M_\pi^3 ~ \widetilde{V}\left( \tau_\pi; R_{dual}, \frac{E_{dual}}{M_\pi}, \frac{M_\rho}{M_\pi}, g_{\rho \pi \pi} \right) ~ , ~
\ee
where $\tau_\pi \equiv M_\pi t$ is the ``pion time'' and the function $\widetilde{V}$ depends only on dimensionless quantities, whose values do not require the knowledge of the absolute scale setting.
Using Eqs.~(\ref{eq:aell_t}-\ref{eq:kernel_ell}) one gets at the physical point
\be
    a_\ell^{\rm HVP, LO}(ud) = 4 \alpha_{em}^2 \int_0^\infty d\tau_\pi ~ \widetilde{K}_\ell( \tau_\pi) ~ 
                                                \widetilde{V}\left[ \tau_\pi; R_{dual}^{phys}, \left( \frac{E_{dual}}{M_\pi} \right)^{phys}, \left( \frac{M_\rho}{M_\pi} \right)^{phys}, 
                                                g_{\rho \pi \pi}^{phys} \right] ~ 
\ee
with
\be
    \widetilde{K}_\ell(\tau_\pi)  = \tau_\pi^2 \int_0^1 dx ~ (1-x) \left[ 1 - j_0^2\left( \frac{m_\ell}{M_\pi^{phys}} \frac{\tau_\pi}{2} \frac{x}{\sqrt{1-x}}\right) \right] ~ . ~
\ee
As described in Appendix~\ref{sec:appA}, in our hadronic scheme we adopt the value $M_\pi^{phys} = M_\pi^{(0),\textrm{FLAG}} = 134.98$ MeV.

\end{document}